\documentclass[journal]{IEEEtran}
\usepackage{amsmath,amsfonts}
\usepackage{algorithmic}
\usepackage{algorithm}
\usepackage{array}
\usepackage[caption=false,font=normalsize,labelfont=sf,textfont=sf]{subfig}
\usepackage{textcomp}
\usepackage{stfloats}
\usepackage{url}
\usepackage{verbatim}
\usepackage{graphicx}
\usepackage{cite}
\usepackage{hyperref}
\usepackage{caption}

\begin{document}
\title{Remote UGV Control via Practical Wireless Channels: A Model Predictive Control Approach}
\author{Jinghao Cao, Subhan Khan, Wanchun Liu, Yonghui Li, Branka Vucetic
\thanks{J. Cao, S. Khan, W. Liu, Y. Li, and B. Vucetic are with the School of Electrical and
Computer Engineering, The University of Sydney, Sydney, NSW 2006,
Australia (e-mail: jinghao.cao@sydney.edu.au; subhan.khan@sydney.edu.au; wanchun.liu@sydney.edu.au; yonghui.li@sydney.edu.au; branka.vucetic@sydney.edu.au)}}

\maketitle

\begin{abstract}
In addressing wireless networked control systems (WNCS) subject to unexpected packet loss and uncertainties, this paper presents a practical Model Predictive Control (MPC) based control scheme with considerations of of packet dropouts, latency, process noise and measurement noise. A discussion of the quasi-static Rayleigh fading channel is presented herein to enhance the realism of the underlying assumption in a real-world context. To achieve a desirable performance, the proposed control scheme leverages the predictive capabilities of a direct multiple shooting MPC, employs a compensation strategy to mitigate the impact of wireless channel imperfections. Instead of feeding noisy measurements into the MPC, we employ an Extended Kalman Filter (EKF) to mitigate the influence of measurement noise and process disturbances. Finally, we implement the proposed MPC algorithm on a simulated Unmanned Ground Vehicle (UGV) and conduct a series of experiments to evaluate the performance of our control scheme across various scenarios. Through our simulation results and comparative analyses, we have substantiated the effectiveness and improvements brought about by our approach through the utilization of multiple metrics. 
\end{abstract}


\section{Introduction}
\label{sec:introduction}
\IEEEPARstart{R}{ecent} advancements in wireless communication networks, sensing technologies, and edge computing have opened up numerous opportunities within domains such as the Internet of Things (IoT). Wireless networked control systems (WNCS) represent a prominent research area within this domain \cite{8166737}. Unlike traditional networked control systems, WNCS employs wireless networks for connectivity instead of wired cable connections. In a WNCS, controllers, sensors, and actuators are distributed spatially. This feature offers distinct advantages, including the provision of additional space for functional expansions and the reduction of infrastructure complexity \cite{7466847,blaney2009wireless,chen2010distributed}. 

Remote Unmanned Ground Vehicle (UGV) control is a practical implementation of WNCS, it has vast applications across military, industrial, and civilian domains. In military applications, UGV can perform reconnaissance, surveillance, and threat neutralization tasks in hostile or inaccessible areas, reducing risks to human life. Industrially, UGVs enhance efficiency and safety in tasks like material handling, inspection, and maintenance in hazardous environments. For civilian uses, UGVs contribute to public safety, emergency response, and even in autonomous transportation systems, showcasing their versatility and potential to revolutionize various sectors by extending human capabilities and safeguarding human lives in dangerous scenarios \cite{gage1995ugv,ni2021review,mohamed2018advanced}.

\subsection{Related Works}

Numerous studies have established the theoretical foundations of remote UGV control from various perspectives. These include stability analysis of adaptive control schemes under dynamic network conditions (such as latency, data rates, and packet irregularities), computational resource allocation of remote control under edge computing context, and dynamic service migration scheme \cite{QUEVEDO20133458,10136631,10197647,9134368,8400587,8016573,9301260}.

To develop a practical and robust remote control model, a thorough investigation into various controller designs across diverse wireless communication environments is needed. Among the numerous existing controller designs, three are considered as the most widely used: the Proportional-Integral-Derivative (PID) controller, the Linear Quadratic Regulator (LQR) control, and the Model Predictive Control (MPC). It is noteworthy that the predictive capability inherent to MPC grants it the possibility to compensate for missing measurement data or input commands, as highlighted in \cite{7098363, mao2020robust}, this may count as one of the advantage of MPC in remote UGV control. 

As a straightforward and effective control method, the PID controller is extensively utilized in WNCS including remote UGV control. In \cite{vasquez2018new,8104807}, four augmented PID control algorithms are introduced to mitigate the impact of continuous packet dropouts in wireless communication networks within control systems. \cite{hussien2020effect} explores the influence of various parameters, such as maximum overshoot, sampling frequency, and latency. 

The LQR controller is also widely recognized in the field of WNCS. As an optimal control strategy, LQR offers analytical solutions that can be computationally advantageous for certain tasks. In \cite{nguyen2015packet} and \cite{yi2017compensation}, multiple compensator designs are proposed, including a predictive compensator, a compensator based on the LQR, and a combination of the previously mentioned compensators. These compensator designs aim to mitigate the effects of packet loss induced by the wireless network.

Due to the predictive nature of MPC, it has been utilized across various research fields, providing robust and well- established solutions for a wide range of scenarios including WNCS. MPC could enhance trajectory and state estimation in autonomous systems, exemplified by its application in vehicles and UGVs through frameworks like CasADi and methodologies including nonlinear and robust MPC. It has shown promise in dynamic environments, utilizing approaches like EFK and SDP-based HMPC for precise navigation \cite{casadi,9491040,kayacan2015robust,khan2022design,10227601,khan2022fast}. Under imperfect wireless connections, MPC could help address challenges like disturbance minimization, packet loss, and network constraints through innovative strategies such as Economic MPC, enhanced tracking control, and min-max optimization, improving reliability and performance \cite{mao2020robust,7501536,li2013networked,jurado2015stochastic}.

\subsection{Contributions}
In the context of real-world industry, MPC and its variants have been widely applied across various scenarios. They have demonstrated robust performance in addressing challenges such as uncertainties and dynamically generated reference trajectories. However, the majority of these studies have operated under the assumption of perfect communication channels, thereby neglecting the investigation of these tasks within an imperfect communication environment. Furthermore, many of these investigations tend to oversimplify the communication channel models or system dynamics.

In contrast to the existing researches (refer to Table \ref{comparetable}), this paper comprehensively addresses numerous uncertainties arising from network imperfections, processing disturbances, and measurement noise affecting our UGV. To mitigate these uncertainties, we propose a control scheme designed to minimize their impacts and ensure the optimal performance of the UGV. Lastly, we conduct multiple simulations to assess the effectiveness of the proposed methodology and to compare it with the existing methodologies. The primary contributions of this paper are outlined as follows:
\begin{enumerate}
    \item We propose a novel approach incorprating the predictive feature of MPC with a delay and packet loss compensation strategy for UGV navigation to accomplish reference tracking tasks. This method takes into account constraints posed by both moving and stationary obstacles, and it has not been previously discussed in the open literature. 
    \item We investigate the impact of both dynamic and static obstacles within the framework of WNCS. In order to evaluate the performance of our control scheme, we have taken into account various metrics, including the sampling period, controller processing time, and the prediction horizon of the MPC, for comparison with existing methodologies.
    \item We integrated the EFK to address process disturbances and measurement noise in the localisation, thereby enhancing the remote UGV control scheme. This integration serves to reinforce the robustness and effectiveness of our proposed method in more complex tasks or environments.
\end{enumerate}
\subsection{Outline}
This paper contains the following sections: The problem formulation is in Section \ref{problemformulation}, it contains the kinematic model, the measurement model, and the wireless communication channel. The controller design is described in Section \ref{controllerdesign}, which includes the MPC control algorithm and the compensation strategy of the packet loss and delay. Section \ref{extendedkalmanfilter} describes the EFK design in the localisation. Section \ref{resultsanddiscussion} presents the outcomes of our simulation experiments and comparative analyses with existing methodologies. And lastly, the conclusions of this work is included in Section \ref{conclusion}.
\begin{table*}[t]
	\renewcommand\arraystretch{1}
	\centering
	\caption{COMPARATIVE TABLE OF VARIOUS PAPERS AND METHODOLOGIES}
	\begin{tabular}{|p{1.5cm}|p{1.5cm}|p{1.5cm}|p{1.5cm}|p{1.5cm}|p{1.5cm}|p{1.5cm}|p{1.5cm}|p{1.5cm}|p{1.5cm}|}
		\hline
		Method & Loss & Delay & Sampling Period & Processing noise & Measurement noise & Controller processing time\\
        \hline
        This paper & \checkmark & \checkmark & \checkmark & \checkmark & \checkmark & \checkmark\\
		\hline
         PID \cite{8104807}& \checkmark & \checkmark & $\times$ & \checkmark & $\times$ & $\times$\\
        \hline
         LQR \cite{nguyen2015packet}& \checkmark & \checkmark & $\times$ & $\times$ & $\times$ & $\times$\\
        \hline
         MPC \cite{mao2020robust}& $\times$ & \checkmark & $\times$ & \checkmark & $\times$ & $\times$\\
        \hline
         MPC \cite{7501536}& \checkmark & \checkmark & \checkmark & $\times$ & $\times$ & $\times$\\
         \hline
         MPC \cite{li2013networked}& \checkmark & \checkmark & \checkmark & $\times$ & $\times$ & $\times$\\
        \hline
	\end{tabular}\\
	\label{comparetable}
\end{table*}
\section{Problem formulation}
\label{problemformulation}
\begin{figure*}[htbp]
\centerline{\includegraphics[width=0.8 \textwidth]{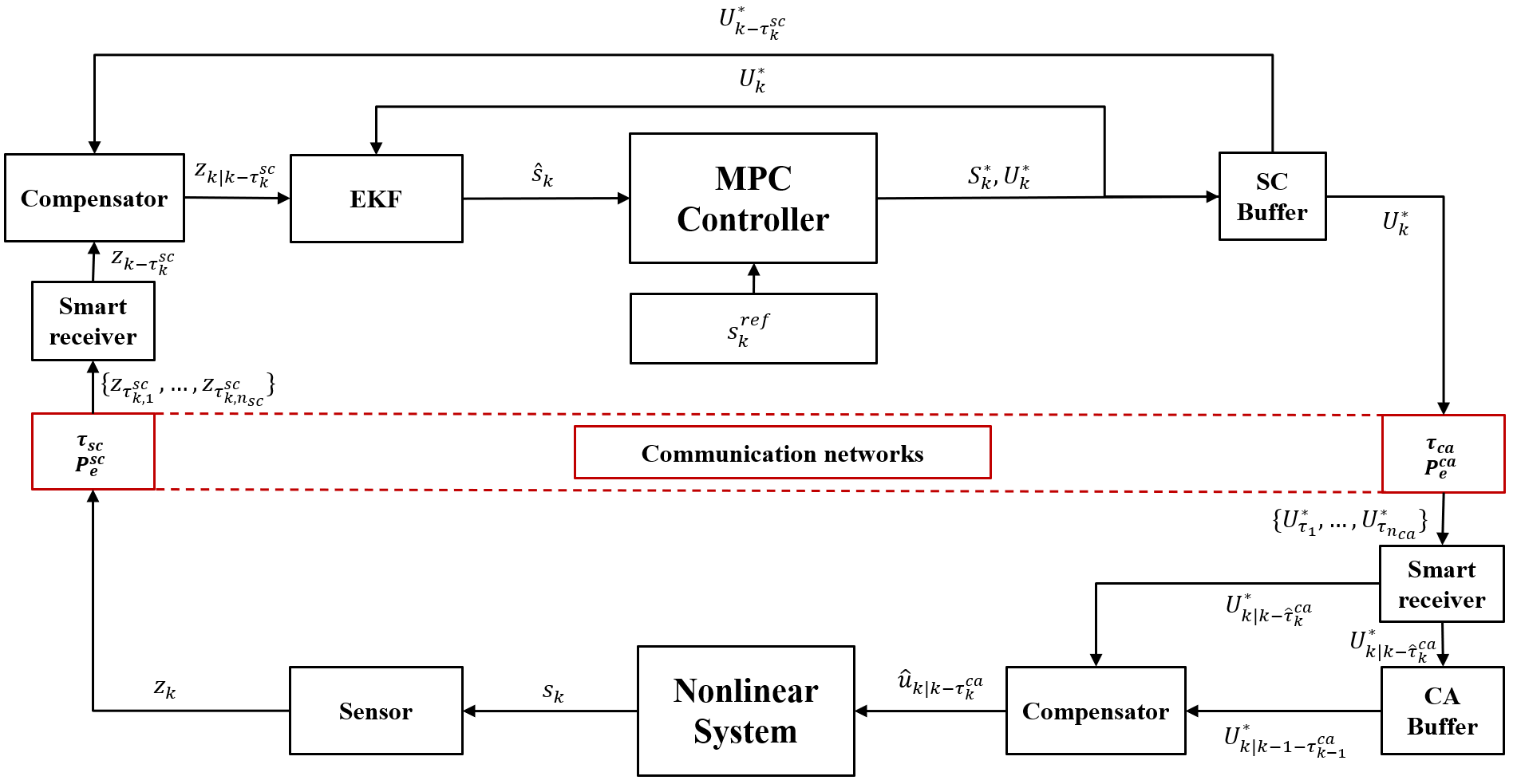}}
\caption{The schematic diagram of the proposed WNCS with communication imperfections illustrated in red.}
\label{block diagram}
\end{figure*}

The kinematic models employed in this paper account for measurement noise caused by imperfect radar readings, control disturbances arising from skidding or slippage, as well as packet loss and transmission latency introduced by the wireless communication channel. The the overall structure of the networked MPC system is shown in Figure \ref{block diagram}. 

The design of the proposed wireless network control system leverages the predictive capabilities of MPC. It employs predicted measurements and control commands for the controller and the dynamic system, respectively, in the presence of network latency or packet loss. Furthermore, this design accommodates noisy measurements, and an EFK is employed to mitigate noise in the localization component of the system. From Table \ref{comparetable}, it is evident that few research papers consider all four discrepancies in the simulation (process noise, measurement noise, packet loss, and communication delay). 

\subsection{UGV kinematic model}
To develop a suitable UGV platform to support our experiments, we derived the following kinematic model from \cite{4509451}.

\begin{align}
\label{eq:cont_dyn} 
    \dot{s}(t) &= f(s(t),u(t)),\\
    \begin{bmatrix}
    \dot{x} (t) \\
    \dot{y} (t)\\
    \dot{\theta} (t)
    \end{bmatrix}
    &=
    \begin{bmatrix}
    v (t) cos(\theta(t)) \\
    v (t) sin(\theta(t)) \\
    \omega (t)
    \end{bmatrix}.
\end{align}

In this kinematic model, there are three states: $s = \left[x,y,\theta \right]^T$  representing x, y-axis positions, and robot orientation \(\theta\). The control input, $u = \left[v,\omega \right]^T$ , represents the speed and angular speed of the robot respectively. Our UGV model is governed by its linear velocity $v(t)$ and angular velocity $\omega(t)$, which are regarded as control input. The agent's position and orientation evolve based on these velocities according to the kinematic model provided. 

Since we are applying control inputs to a discrete-time system, this paper employs a first-order Euler discretization process. Let $\Delta T$ represent the sampling time. The discrete-time dynamics can be expressed as follows:
\begin{align}
    s_{k+1} &= f(s_k,u_k),\\
    \begin{bmatrix}
    x_{k+1} \\
    y_{k+1}\\
    \theta_{k+1}
    \end{bmatrix}
    &= 
    \begin{bmatrix}
    x_k \\
    y_k\\
    \theta_k
    \end{bmatrix}
    +
    \Delta T
    \begin{bmatrix}
    v_k cos(\theta_k) \\
    v_k sin(\theta_k) \\
    \omega (k)
    \end{bmatrix}.
\end{align}
With the presence of disturbances, the kinematic model becomes:
\begin{align}
\label{dynamic_disturbance}
    \begin{bmatrix}
    x_{k+1} \\
    y_{k+1}\\
    \theta_{k+1}
    \end{bmatrix}
    &= 
    \begin{bmatrix}
    x_k \\
    y_k\\
    \theta_k
    \end{bmatrix}
    +
    \Delta T
    \begin{bmatrix}
    \bar{v}_k cos(\theta_k) \\
    \bar{v}_k sin(\theta_k) \\
    \bar{\omega}_k
    \end{bmatrix}.
\end{align}
where $\bar{v} = v + d_v$ and $\bar{\omega} = \omega + d_\omega$ are disturbed control inputs. $d_v$ and $d_\omega$ are additive Gaussian noise with standard deviations $\sigma_v$ and $\sigma_\omega$.

\subsection{Mobile robot measurement model}
In designing this problem, we will use a single sensor and one access point (AP), both positioned at the origin (which will also be the starting point of the robot). The AP, or Access Point, is a device that facilitates the connection of wireless devices to a wired network through Wi-Fi or comparable standards, thereby serving as a bridge or intermediary.
The sensor depicted in Figure \ref{block diagram} is identified as a sensing beacon capable of capturing state information of the target UGV of interest. The motion of the UGV will be periodically detected by the sensor, after which the state information will be updated and transmitted to a centralized computer through a wireless network.
In our design, the data obtained consists solely of the relative distance from the beacon, represented by $r$, and the relative bearing angles with respect to the robot, denoted by $\alpha$. The measurement model can be formulated as follows:
\begin{align}
    z_k &= h(s_k) + n_y,\\
    \begin{bmatrix}
    r_k \\
    \alpha_k
    \end{bmatrix}
    &= 
    \begin{bmatrix}
    \sqrt{x_k^2 + y_k^2} \\
    \arctan\left(\frac{y_k}{x_k}\right) 
    \end{bmatrix}
    +
    \begin{bmatrix}
    n_r \\
    n_\alpha 
    \end{bmatrix}.
\end{align}
where $n_r$ and $n_\alpha$ are additive Gaussian noise with standard deviations $\sigma_r$ and $\sigma_\alpha$.

\subsection{Wireless Communication Channel}

The presence of communications introduces imperfections to the control systems. In this investigation, we shall delve into the impacts of random delays and packet losses in networked control systems. Our objective is to characterize these imperfections utilizing a Quasi-Static Rayleigh Fading Packet Loss Model and a set of mutually independent stochastic delay models.

\subsubsection{Communication Latency}

In our design of WNCS, we incorporate stochastic delays within both sensor-to-controller and controller-to-actuator communication channels. We adopt an independent stochastic delay model, which posits that these delays are to be considered as independent random variables. Each delay is characterized by a distinct stochastic function, as proposed in the model by \cite{ge2013modeling}. To quantify the variability of the random delay within our framework, we introduce a maximum latency threshold, $\tau_{max}$. This threshold facilitates the modeling of random delay as a discrete uniform distribution, described by the following probability function:
\begin{align*}
P(\tau_{k} = d) = \frac{1}{\tau_{max} + 1}.
\end{align*}
where $\tau_k$ denotes the random delay experienced at the $k$-th time step. Here, $d$ is an integer within the inclusive range of $0$ to $\tau_{max}$, signifying that each potential delay duration has an equal probability of occurrence.

\subsubsection{Packet loss model}
In the design of our WNCS, we postulate the existence of quasi-static Rayleigh fading channels. This assumption implies that the channel conditions conform to the Rayleigh fading model throughout the duration of each block transmission. The relationship between the received signal ${r_k}'$ and the transmitted signal $r_k$ at time step $k$ considering unit transmit power is given by:
\begin{align}
r^{'}_k = \beta_k e^{j\phi_k} r_k .
\end{align}

In this expression, $\beta_k$ delineates the amplitude gain, and $\phi_k$ designates the phase shift introduced to the signal at the k-th time step, assuming unit transmit power. These parameters are modeled as stochastic variables: $\beta_k$ is distributed according to the Rayleigh probability density function $f(\beta ; \sigma) = \frac{\beta}{\sigma^2}exp(-\beta^2/2\sigma^2)$ , while $\phi_k$ is uniformly distributed. The scale parameter $\sigma$ reflects the dispersion of the channel gains, with larger values of $\sigma$ indicating a more extensive range of channel gain fluctuations.

The packet error probability $P_e$ for a block with a length $L$ and a coding rate $R$ can be derived as outlined in reference \cite{9145394}.
\begin{align}
    Pe(1) &= P_0(R) + \frac{log(L)}{L}\phi \log(R) + \frac{1}{L}\phi_0(R),\\
    P_0(R) &= 1 - \exp\left(\frac{-e^R -1}{\gamma}\right),\\
    \phi &= \frac{-e^R}{2\gamma} \exp\left(\frac{-e^R-1}{\gamma}\right),\\
    \begin{split}
        \phi_0 &= \frac{e^R}{\gamma}\exp\left(\frac{-e^R -1}{\gamma}\right) \\
        &\quad \times \left(2-\frac{-e^R -1}{\gamma} + \log \left[\frac{1}{\sqrt{2 - e(1-e^{-2R})}}\right]\right).
    \end{split}
\end{align}
In this context, $\gamma$ stands for the signal-to-noise ratio (SNR) associated with the current block transmission. The expected packet error rate can be denoted as $\mathbb{E}_{\gamma}[\text{P}_e(1)]$, where $\text{P}_e(1)$ represents the packet error probability for an individual transmission. The symbol $\mathbb{E}_\gamma$ corresponds to the expected value of the Rayleigh distribution.

\section{Controller design}
\label{controllerdesign}
The controller and the nonlinear systems are physically isolated from each other. Therefore, a networked control protocol becomes essential to effectively manage the end-to-end data exchange process. Within our design framework, wireless communication introduces both latency and packet loss. Consequently, the successful reception of each sent packet within the current time step cannot be guaranteed. For instance, at time step $k$, the measurement $z_k$ is transmitted but due to latency $\tau_k$, it is only received at the controller's end during time step $k+\tau_k$. If there is a packet loss, no data will be received from the sensor.

In the scenario when multiple packets are received at time step $t$, we have designed a smart packet receiving dynamics (smart receiver) that will prioritize the packet with the smallest AoI. The AoI can be obtained by examining the timestamp of each packet. The latency of the transmitted packet (from sensor to controller) can be formulated as follows:
\begin{equation}\label{ranging error}
    \tau^{sc}_k = \left\{
    \begin{array}{lcl}
    \Tilde{\tau^{sc}_k }& \text{, if $k-1-\tau_{k-1} < \Tilde{\tau^{sc}_k }$ ,} \\
    1 +\tau^{sc}_{k-1} & \text{, else.}
    \end{array}
    \right.
\end{equation}
Here, $\Tilde{\tau^{sc}_k } = \min_{\tau}\{T_{k}^{sc}\}$, and $T^{sc}_k := \{\tau^{sc}_{k,1},...,\tau^{sc}_{k,n_k^{sc}}\}$, $n_k^{sc}\in \mathbb{Z}^+$ represents the number of received packets at S-C smart receiver at time step $k$. At the sensor-to-controller end specifically, assume the smart receiver at the controller end receives multiple packets of measurement information with different delays $\{z_{\tau_{k,1}^{sc}},...,z_{\tau_{k,n_k^{sc}}^{sc}}\}$, using the smart receiver mentioned above we select the packet with the smallest delay $z_{k-\tau_k^{sc}}$. But if $z_{k-\tau_k^{sc}}$ is a delayed measurement, directly applying it to the EKF may cause bigger errors, so here we use the optimal control sequences stored in the buffer to estimate the $z_{k|k-\tau^{sc}_k}$ for the current time step $k$, and this part refers to the Compensator in Figure \ref{block diagram}. 

The details of how the compensator works goes as follows. Assume the Sensor-to-Controller buffer stores all the historical records of the MPC outputs, according to the packet with the smallest delay $z_{k-\tau_k^{sc}}$ selected by the smart receiver and the historical control inputs stored in the SC buffer $U^*_{k-\tau^{sc}} = \{u^*_{k-\tau^{sc}},...,u^*_{k-\tau^{sc}+N} \}$, we could predict the state measurements of the time step $k$ based on the delayed measurement $z_{k-\tau^{sc}_k}$, and this predicted state measurement is denoted as $z_{k|k-\tau^{sc}_k}$.

The controller utilized in our design is Multiple Shooting MPC. This form of MPC employs both the system states and control inputs as optimization variables. For a prediction horizon denoted as $N$, the MPC controller computes $N$ successive optimal state values $S_k^*=\{s^*_{k},...,s^*_{k+N}\}$ along with corresponding control inputs $U_k^*=\{u^*_{k},...,u^*_{k+N}\}$. These sequences of predicted states and control inputs are then respectively stored in the Sensor-to-Controller buffer and the Controller-to-Actuator buffer.

Consequently, at time step $k$, with the delayed measurement $z_{k-\tau^{sc}_k}$, the estimated states generated by the EFK is represented as $\hat{s}_{k}$, while the control input sent to the actuator is denoted as $\hat{u}_{k}$. As a result, for the dynmaic system given in equation \ref{dynamic_disturbance}, the multiple shooting MPC problem could be solved by finding out the optimal state $z$ and $u$ for the following optimization problem:
\begin{align}
\label{MPC_opt}
\ell(s_k,u_k) &= ||s_k-s^{ref}_k||^2_Q + ||u_k-u^{ref}_k||^2_R,\\
\min_{s,u}\ \ \   &\ell_{k+N}(\hat{s}_{k+N},\hat{u}_{k+N}) + \sum_{i=0}^{N-1} \ell_k(\hat{s}_{k+i},\hat{u}_{k+i}),\\
\begin{split}
    s.t. \ \ \ &\hat{s}_{k+1} = f(\hat{s}_{k},\hat{u}_{k}),\\
    &r_{rob} + r_o \leq ||(\hat{x}_{k+i},\hat{y}_{k+i|k-\tau^{sc}_k})-(x_{o,j},y_{o,j})||,\\
    &|v_{k}^{max}| \leq V_{max}, \\
    &|\omega_{k}^{max}| \leq \Omega_{max},\\
    &\forall k \in \{k,...,k+N\},\\
    &\forall j \in \{1,...,n_o\}.
\end{split}
\end{align}
In the given context, $N$ represents the length of the prediction horizon, $V_{max}$ signifies the robot's maximum achievable velocity, $\Omega_{max}$ denotes the maximum achievable angular velocity of the robot, and $n_o$ corresponds to the count of obstacles necessitating to be avoided.

At time step $k$, we provided the pseudo-code of the algorithm in the Algorithm \ref{controller protocal}.
\begin{algorithm}
\caption{WNCS design}
\begin{algorithmic} [1]
\label{controller protocal}

\STATE \textbf{Sensor to controller:}
\STATE Find $\Tilde{\tau^{sc}_k } = \min_{\tau}\{T_{k}^{sc}\}$
\IF{$k-1-\tau^{sc}_{k-1} < \Tilde{\tau^{sc}_k }$}
\STATE $\tau^{sc}_k = \Tilde{\tau^{sc}_k }$
\ELSE
\STATE $\tau^{sc}_k = 1 + \tau^{sc}_{k-1}$
\ENDIF
\STATE Estimate $z_{k|k-\tau^{sc}}$ based on $z_{k-\tau^{sc}}$ and $U^*_{k-\tau^{sc}}$ (historical control sequence stored in buffer)
\STATE Feed $z_{k|k-\tau^{sc}_k}$ into the EKF, and get the state estimation $\hat{s}_{k|k-\tau^{sc}_k}$
\STATE Use $\hat{s}_{k|k-\tau^{sc}_k}$ as the initial condition of (\ref{MPC_opt})
\STATE Solve equation (\ref{MPC_opt}), and store the optimal solutions $S_k^*:=\{s^*_{k},...,s^*_{k+N}\}$ and $U_k^*:=\{u^*_{k},...,u^*_{k+N}\}$
\STATE Store $Z_k^*$ in the Sensor-to-Controller buffer, send $U_k^*$ to the Actuator.

\STATE \textbf{Controller to actuator:}
\STATE Find the control input packet with the smallest AoI $\Tilde{\tau^{ca}_k } = \min_{\tau}\{T_{k}^{ca}\}$ 
\IF{$k-1-\tau^{ca}_{k-1} < \Tilde{\tau^{ca}_k }$}
\STATE $\tau^{ca}_k = \Tilde{\tau^{ca}_k }$
\ELSE
\STATE $\tau^{ca}_k = 1 + \tau^{ca}_{k-1}$
\ENDIF
\STATE Store $U^*_{k-\tau^{ca}_k}$ in the Controller-to-Actuator buffer
\STATE Apply $\hat{u}_{k|k-\tau^{ca}_k}$ to the system

\end{algorithmic}
\end{algorithm}

\section{Extended Kalman filter}
\label{extendedkalmanfilter}
In real life, we could only get access to noisy measurements instead of actual position information. We used an EFK in this part to update the state information of our agent. 

To obtain the optimal estimate of $\hat{z}$, the updating mechanism of the Kalman filter can be derived as follows:
\begin{align}
    \hat{s}_{k+1|k} &= f(\hat{s}_{k},\hat{u}_k),\\
    \hat{s}_{k+1|k+1} &= \hat{s}_{k+1|k} + K_{k+1}(z_{k+1} - h(\hat{s}_{k+1|k})).
\end{align}
In the given context, $K_k$ represents the nearly optimal Kalman gain, $\hat{u}_k$ is the received control input at time step $k$. Let's consider a beacon with an observation sequence $y = {y_1, y_2, ..., y_k}$ until time $k$. Based on these observations, the initial state estimate derived from $y_k$ is denoted as $\hat{z}_{k|k}$.

When we have the current state estimate $\hat{z}_{k|k}$ available, we can create linear approximations by introducing Jacobian matrices.
\begin{align}
    A_k &= \frac{\partial}{\partial z} f(s,u)\bigg |_{z = \hat{s}_{k|k}, u = u_k},\\
    B_k &= \frac{\partial}{\partial d} f(s,u)\bigg |_{z = \hat{s}_{k|k}, u = u_k},\\
    C_k &= \frac{\partial}{\partial z} h(s)\bigg |_{z = \hat{s}_{k|k-1}},\\
    D_k &= \frac{\partial}{\partial n} h(s)\bigg |_{z = \hat{s}_{k|k-1}}.
\end{align}
In the EKF's update strategy, the Kalman gain matrix $K$ and the covariance estimation matrix $P$ can be obtained by:
\begin{align}
    P_{k+1|k} &= A_{k+1} P_{k|k} A^T_{k+1} + B_{k+1} Q_{k+1} B_{k+1}^T,\\
    S_{k+1} &= C_{k+1} P_{k+1|k} C_{k+1}^T + R_{k+1},\\
    K_{k+1} &= P_{k+1|k}C^T_{k+1}S^{-1},\\
    P_{k+1|k+1} &= (I-K_{k+1}H_{k+1})P_{k+1|k}.
\end{align}
At time step $k+1$, the measurement $z_{k+1}$ will be fed into the filter, and the state prediction $\hat{s}_{k+1}$ will be used as the input state for the controller.

\section{Results and discussions}
\label{resultsanddiscussion}
\subsection{Overview}
\label{resultsanddiscussionoverview}
In this section, a series of simulation experiments have been conducted to evaluate the applicability of the aforementioned method across a range of tasks. In conducting these experiments, we incorporate a range of sampling intervals, prediction horizons, MPC processing times, and diverse communication channel conditions, including latency, packet error rate, and signal-to-noise ratio. The proposed methodology will be utilized to address tasks related to point stabilization, circular curve tracking, and eight-curve tracking problems. 

\begin{figure}[htbp]
\centerline{\includegraphics[width=0.6 \textwidth]{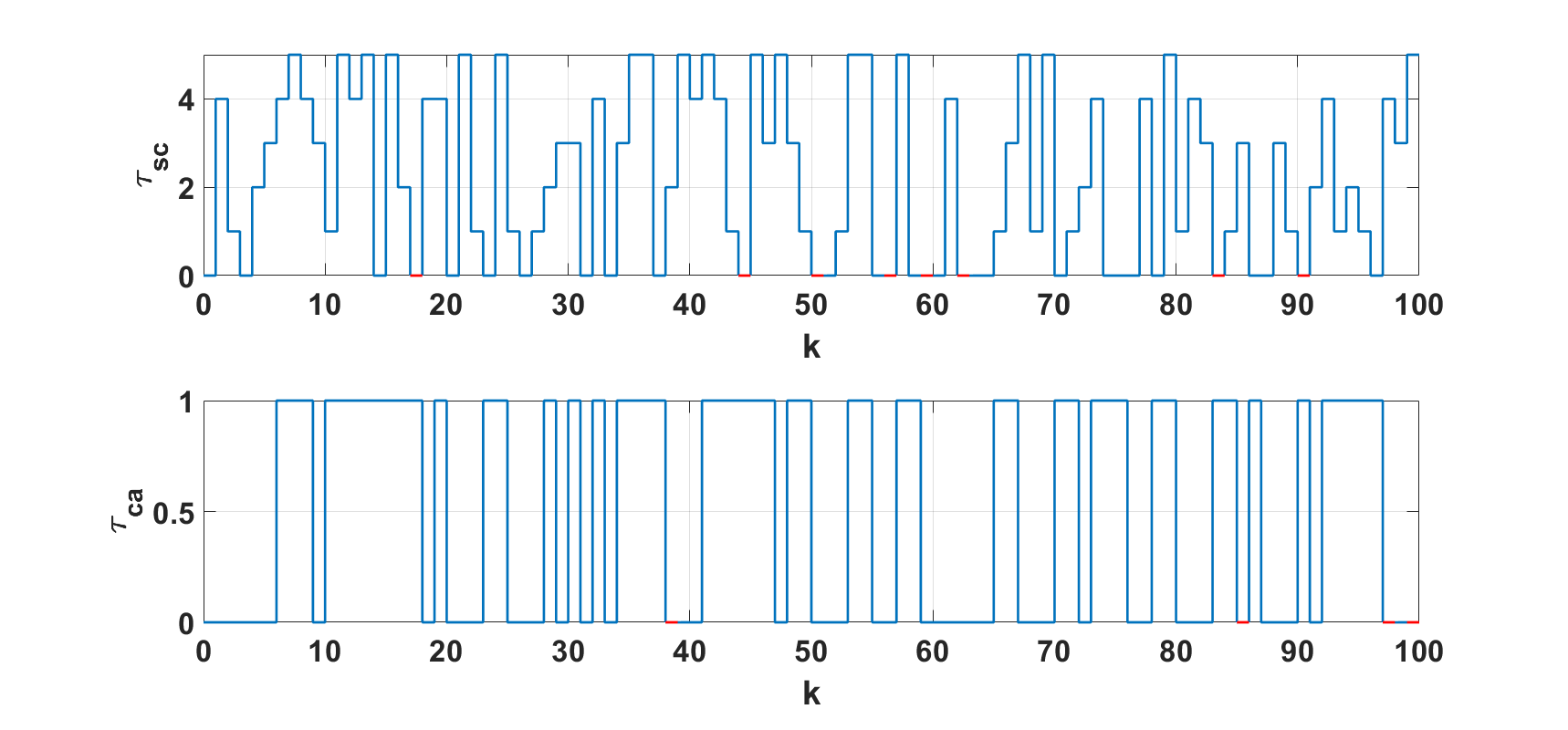}}
\caption{Delay and packet loss sequences of the S-C and C-A channels}
\label{delaysequence}
\end{figure}

\begin{table*}[t]
	\renewcommand\arraystretch{1}
	\centering
	\caption{COMPARATIVE ANALYSIS OF KEY DESIGN STRATEGIES ACROSS DIFFERENT STUDIES}
	\begin{tabular}{|p{1.5cm}|p{2cm}|p{2cm}|p{2cm}|p{2cm}|p{2cm}|p{2cm}|p{2cm}|p{2cm}|p{2cm}|}
		\hline
		Method & Delay mitigation design& Loss mitigation design& Predictive design & Processing noise mitigation design& Measurement noise mitigation design& Obstacle avoidance ability\\
        \hline
        This paper & \checkmark & \checkmark & \checkmark & \checkmark & \checkmark & \checkmark\\
		\hline
         PID \cite{8104807}& $\times$ & \checkmark & \checkmark & \checkmark & $\times$ & $\times$\\
        \hline
         LQR \cite{nguyen2015packet}& \checkmark & \checkmark & \checkmark & $\times$ & $\times$ & $\times$\\
        \hline
         MPC \cite{mao2020robust}& $\times$ & \checkmark & \checkmark & \checkmark & $\times$ & \checkmark\\
        \hline
	\end{tabular}\\
	\label{comparetable4methods}
\end{table*}
In our simulations, the principal parameters are set as follows: the process noise covariance matrix is defined as $Q = \text{diag}([\sigma_v^2, \sigma_w^2]) = \text{diag}([0.005, 0.0349])^2$, and the measurement noise covariance matrix is defined as $R = \text{diag}([\sigma_r^2, \sigma_\alpha^2]) = \text{diag}([0.1, 0.0349])^2$. The prediction horizon and the sampling time are set to $N = 100$ and $T_s = 10 \text{ms}$ respectively. 
The nominal sensor-to-controller (S-to-C) communication channel condition is configured with $\text{SNR} = 15$ dB and a maximum delay of $50$ ms. Similarly, the nominal controller-to-actuator (C-to-A) communication channel condition is configured with $\text{SNR} = 20$ dB and a maximum delay of $10$ ms. For both these channels, the code rate is set to be 1, and the block length is defined to be 100 bits. Figure \ref{delaysequence} illustrates the delay and packet loss sequences for the initial 100 time steps of both the S-C and C-A channels. The vertical axis is scaled in terms of the minimum time slot duration ($10 \text{ms}$), while the horizontal axis represents the time steps. Packet loss is indicated in red on the figure to signify transmission failures occurring at specific time steps.
Additionally, the control policies are generated while considering constraints imposed by both moving and stationary obstacles. Given the presence of process noise and measurement noise, an additional safety margin is necessary to ensure the robot's safe task completion. Therefore, instead of employing a robot radius of $0.34$ meters, we applied a safety margin of $0.54$ meters to the MPC.

In order to enhance the fidelity of our simulation and align it more closely with real-life scenarios, we incorporated the following physical specifications into the MPC framework:
\begin{itemize}
    \item UGV Radius: 34 cm
    \item  Safety Margin: 54 cm
    \item Maximum Forward/Backward Speed: 1 m/s
    \item Rotation Speed: 140°/s
\end{itemize}
Notably, we only focus on Quasi-Static Rayleigh Fading channels, thus the channel model variations are not considered in this study.

To assess the performance of our UGV and its associated algorithm in comparison to existing methodologies, we conducted three experiments. In experiment 1 and 2, The simulation framework encompasses dynamic and static obstacles, requiring the autonomous agent to simultaneously achieve reference tracking and obstacle avoidance objectives. The experiments are conducted under nominal conditions, necessitating the agent and the control system to address the challenges posed by transmission, actuation, and measurement inaccuracies. Table \ref{comparetable4methods} listed six key designs that required in our experiment: transmission loss mitigation design, delay compensation design, predictive design of controller, measurement noise and process noise mitigation design (associated with localization and actuation), and the capability for obstacle avoidance. To ensure a robust comparison, all experiments were conducted under uniform operational conditions (nominal conditions). Most of the listed methodologies have transmission delay compensation design and predictive design, but the rest designs are not.

In the comparative analysis, a representative control strategy was selected from each of the three extensively employed WNCS methodologies(PID, LQR, and MPC), and from Table \ref{comparetable4methods}, we could see that the method proposed by this paper has considered all six designs, while other methodologies only considered a part of them. The absence of an effective obstacle avoidance mechanism in the LQR and PID is considered as a common disadvantage for these two control schemes. In contrast, the MPC methods by \cite{mao2020robust} only considered package delay but not packet loss, it also ignore the possible measurement noise that might occur in real-time localisation.

Experiment 3 is the point stabilisation problem, and the objective of this experiment is to analyze the impact of internal and external parameters, such as prediction horizon, latency, and packet loss, on the performance of the proposed method. Employing an experiment of such simplicity facilitates the derivation of more definitive and general conclusions.

\subsection{Experiment 1: Circular curve reference tracking}
In this study, we carried out an experiment focusing on circular-curve reference tracking to evaluate the effectiveness of our proposed approach. The experimental conditions were consistent with the nominal settings previously described. Additionally, we took into account the presence of both dynamic and static obstacles. The reference trajectory follows the circular shape follows the equations below:
\begin{align*}
    x_k^{ref} &= 5 \cos(\theta_k),\\
    y_k^{ref} &= 5 \sin(\theta_k).
\end{align*}

For the parameter $k$, where $k \in (0, T_{\text{sim}}/T_s)$, $T_{\text{sim}}$ denotes the total simulation time, and $T_s$ represents the sampling time. In the specific experiment under consideration, we have set $T_{\text{sim}} = 50$ s and $T_s = 10$ ms. The initial conditions for the UGV have been established as $[x_i, y_i, \theta_i] = [5, 0, \pi/2]$. Similar to other experiments within this paper, the sensor is positioned at the origin.

\begin{figure*}
  \centering
  \subfloat[\label{cir1a}]{%
    \includegraphics[width=0.3\linewidth]{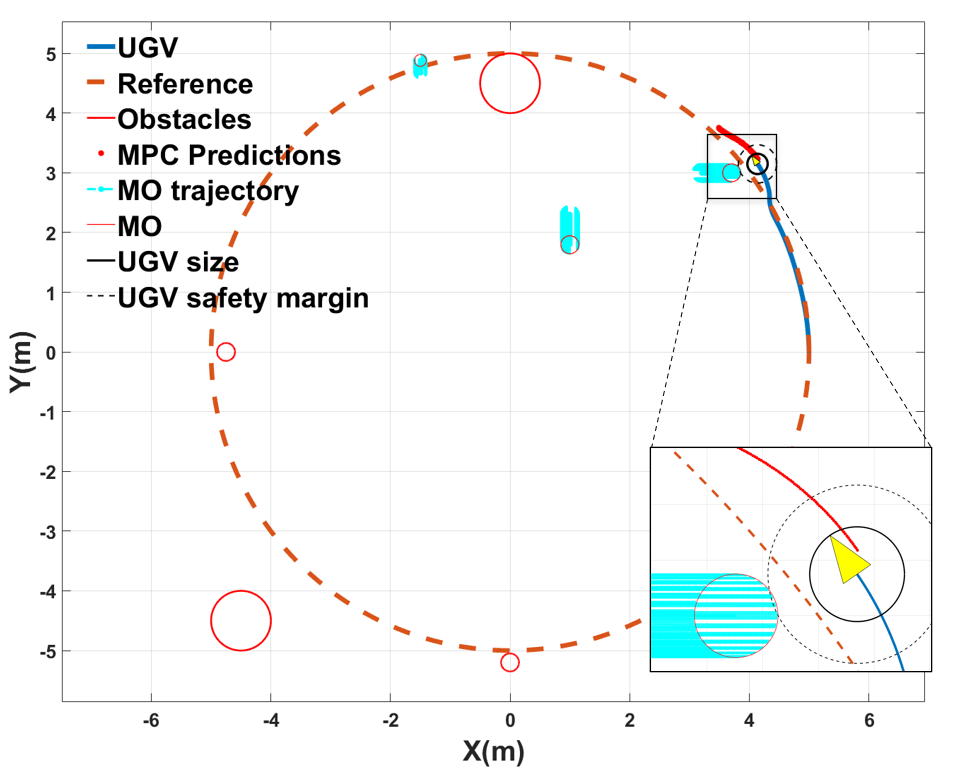}}
  \hfill
  \subfloat[\label{cir1b}]{%
    \includegraphics[width=0.3\linewidth]{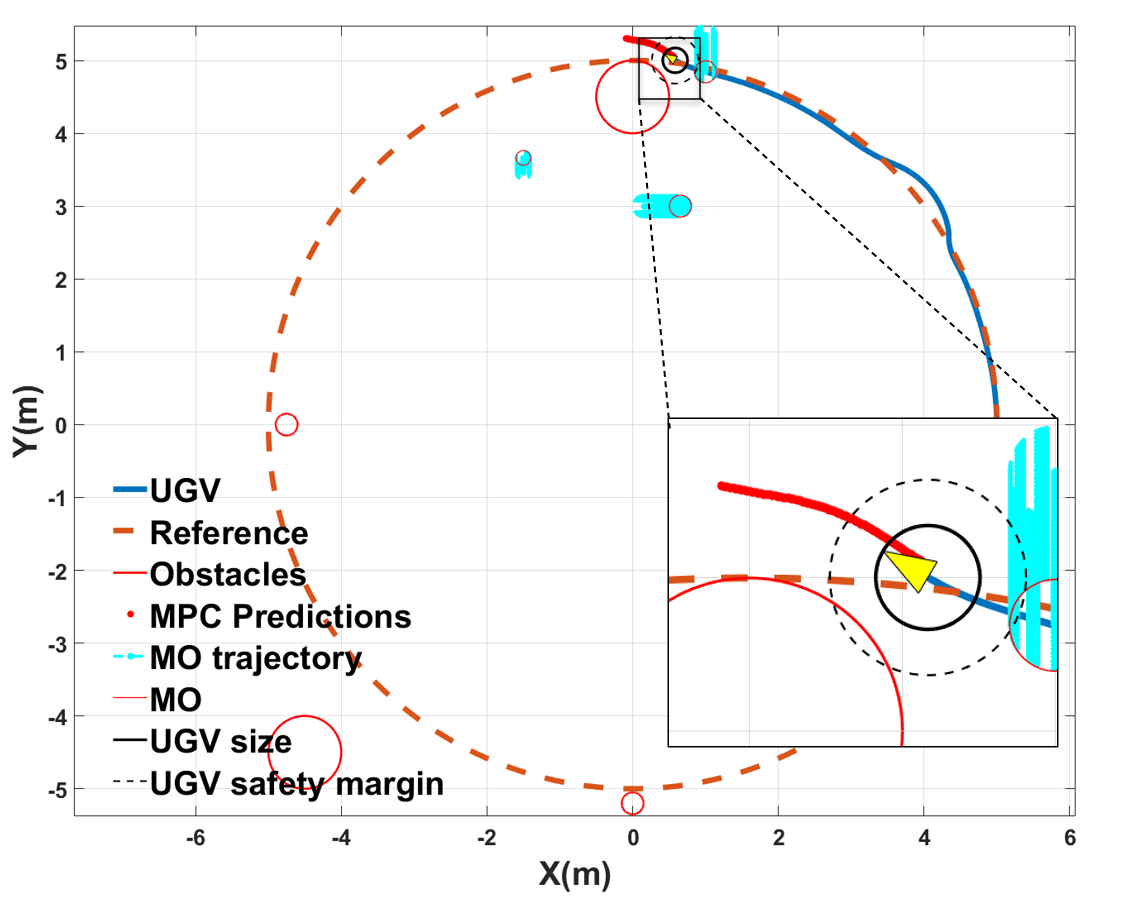}}
  \hfill
  \subfloat[\label{cir1c}]{%
    \includegraphics[width=0.3\linewidth]{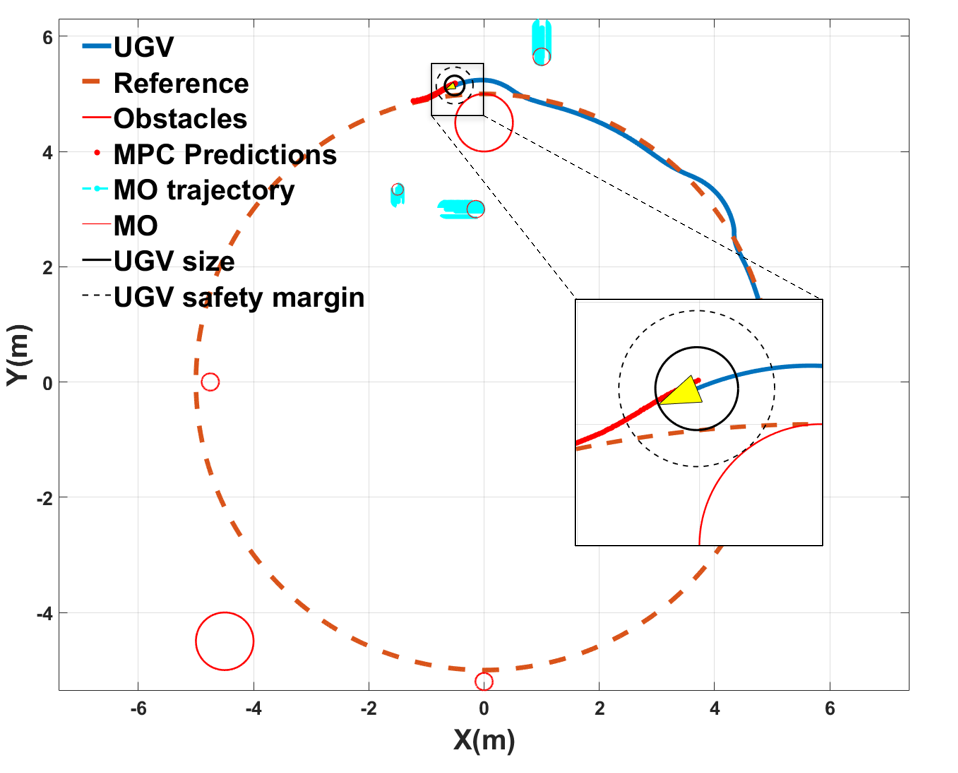}}
  \\
  \subfloat[\label{cir1d}]{%
    \includegraphics[width=0.3\linewidth]{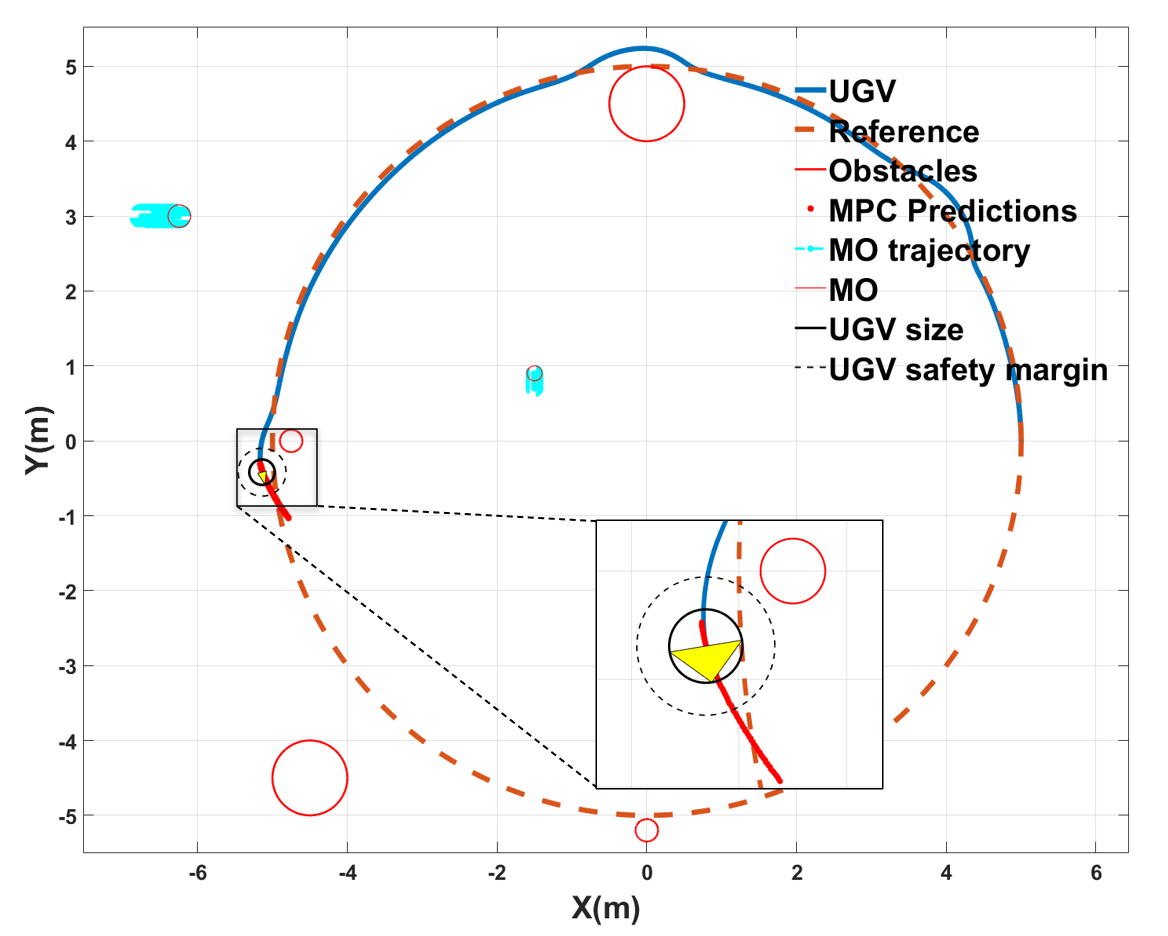}}
  \hfill
  \subfloat[\label{cir1e}]{%
    \includegraphics[width=0.3\linewidth]{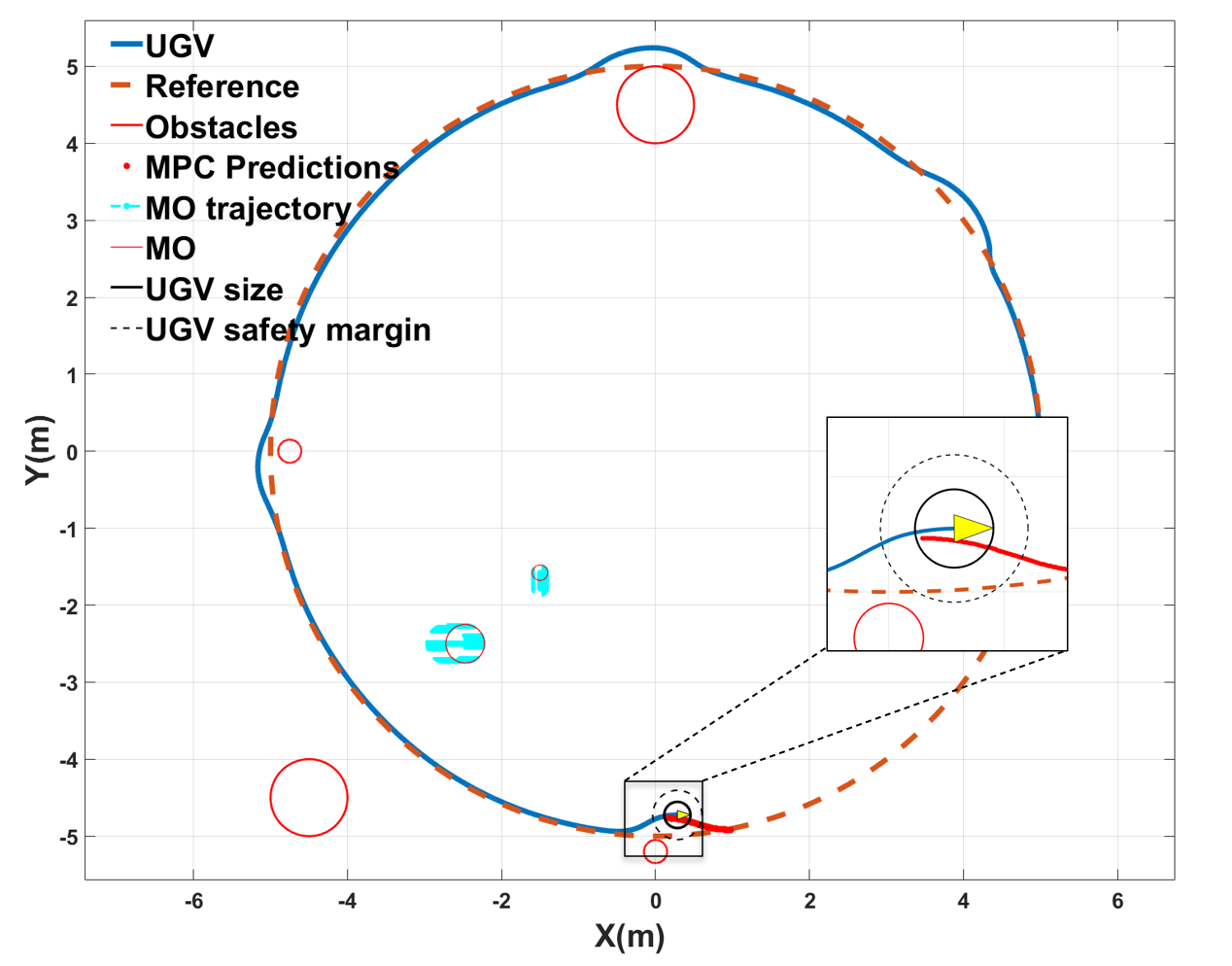}}
  \hfill
  \subfloat[\label{cir1f}]{%
    \includegraphics[width=0.3\linewidth]{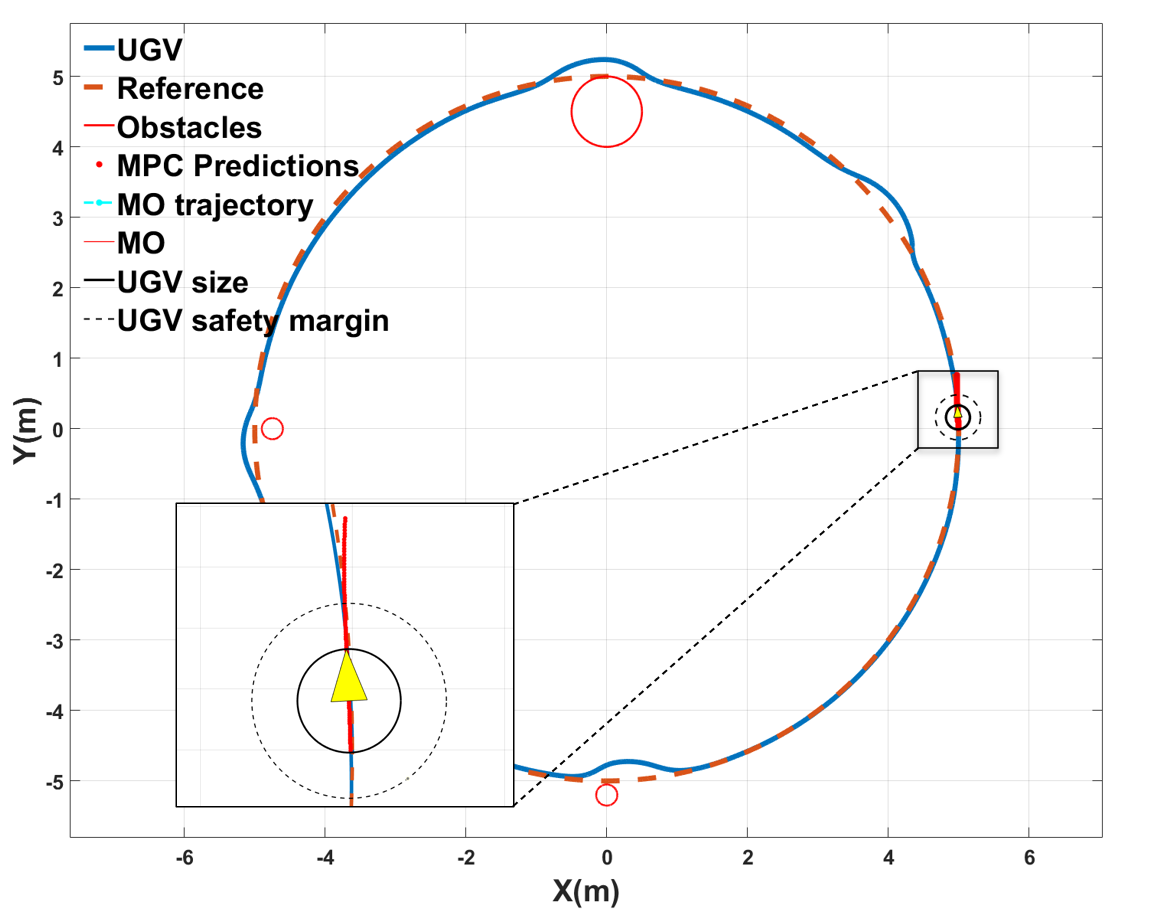}}
  \caption{(a) Avoid the first MO by predicting its motion. (b) Avoid the second MO by fast orientation steering. (c) Avoid the first SO and catching up with the reference. (d), (e) Avoid two more SO and get back to the track. (f) Reach the goal point safely.}
  \label{figcircle}
\end{figure*}

\begin{figure}[htbp]
\centerline{\includegraphics[width=0.35 \textwidth]{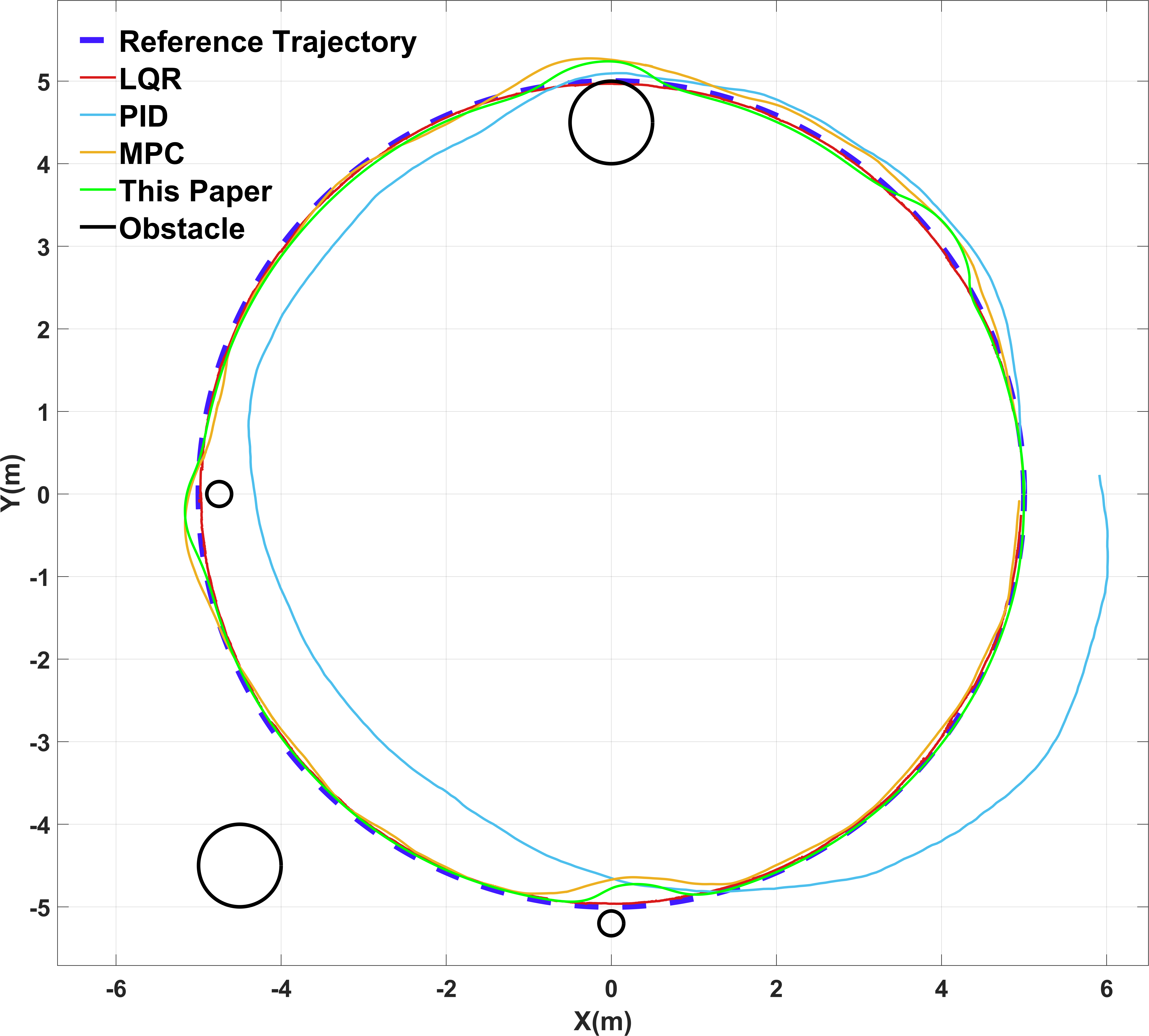}}
\caption{Experiment 1: Comparing of the performance of different methods}
\label{circlecompare}
\end{figure}

In the conducted experiment, we have considered four stationary obstacles (SO) shapes and four moving obstacles (MO), as depicted in Figure \ref{figcircle}. The positions of the stationary obstacles are $(x^{so}_1,y^{so}_1) = (0, 5.2)$, $(x^{so}_2,y^{so}_2) = (0, 4.5)$, $(x^{so}_3,y^{so}_3) = (-4.75, 0)$, and $(x^{so}_4,y^{so}_4) = (-4.5, -4.5)$, each with a respective radius of 0.15m, 0.5m, 0.15m, and 0.5m. And the starting states of moving obstacles are $(x^{mo}_1,y^{mo}_1) = (7,-2.5,\pi)$, $(x^{mo}_2,y^{mo}_2) = (1,-1,\pi/2)$, $(x^{mo}_3,y^{mo}_3) = (-1.5,6,-\pi/2)$, and $(x^{mo}_4,y^{mo}_4) = (6.5,3,-\pi)$ with moving speed of 0.25 m/s, 0.5 m/s, 0.2 m/s, 0.5 m/s, and radius 0.25 m, 0.15 m, 0.1 m and 0.15 m  respectively.


Figure \ref{figcircle} illustrates five typical obstacle avoidance behaviors. In Figure \ref{cir1a}, our UGV would collide with the rear of Mobile Object 4 (MO 4) if it strictly adhered to the circular track. The MPC algorithm computes the optimal avoidance strategy to facilitate a detour. In Figure \ref{cir1b}, if the UGV follows the circular track, there might be two collisions (SO 2 and MO 2). At this stage, the accumulative localisation errors is already not neglectable, and the edge of both obstacles are exceeding the safety region. However, the UGV could still avoid these obstacles and get back to the circular track. In Figure \ref{cir1c} and \ref{cir1d}, it is obvious to see that the y-axis localisation error has reached 10cm, since there is a safety margin for the UGV, all the obstacles could be avoided. In Figure \ref{cir1e}, the localisation error as reached 15cm, the accuracy of the localisation does make the MPC control more difficult, but the UGV could still get back to the circular reference and finish the simulation at $[5,0,\pi/2]$ as shown in Figure \ref{cir1f}.

In comparison with other methods, Fig \ref{circlecompare} shows the behaviours of how agent will deal with this type of task in all four interested methods including LQR from \cite{nguyen2015packet}, PID from \cite{8104807}, MPC from \cite{mao2020robust}. 
Table \ref{eightcomparetableperformance} presents the multi-core processing time, average Euclidean distance error, and step heading error comparisons across different control strategies. Despite the occurrence of average Euclidean distance errors resulting from the implementation of essential obstacle avoidance maneuvers, the approach proposed in this study exhibits remarkable precision in reference tracking, surpassing other methodologies in terms of accuracy. Furthermore, the LQR method demonstrates strong reference tracking capabilities, ranking as the second most effective among all evaluated methods. On the other hand, the MPC method shows proficiency in navigating around obstacles, yet it faces challenges in fulfilling the task within the imposed physical limitations, such as velocity and angular velocity constraints; nevertheless, its performance is deemed satisfactory when considering its obstacle avoidance capabilities. In contrast, the PID control strategy displays the least favorable performance among the evaluated techniques.

In evaluating the computational expense, the PID controller exhibits the minimal processing duration owing to its exceedingly straightforward algorithmic structure. Conversely, the computational demand of our proposed methodology is similar to that of the MPC approach. Meanwhile, the LQR method incurs the most significant average computational overhead, potentially giving rise to additional complications.

\begin{table}[t]
	\renewcommand\arraystretch{1}
	\centering
	\caption{Experiment 1: COMPARATIVE TABLE OF VARIOUS PAPERS AND METHODOLOGIES IN TERMS OF PERFORMANCE}
	\begin{tabular}{|p{1.2cm}|p{1.2cm}|p{1.2cm}|p{1.2cm}|p{1.2cm}|p{1.2cm}|p{1.2cm}|p{1.2cm}|p{1.2cm}|p{1.2cm}|}
		\hline
		Method & Multi-core Process Time (ms) & Average Euclidean distance error(cm) & Step Heading Error(deg)  & Collision Avoidance\\
        \hline
        This paper & 32.47 & 22.38 & 2.214 & \checkmark\\
		\hline
         PID \cite{8104807}& 0.025 & 80.97 & 4.671 & NA\\
        \hline
         LQR \cite{nguyen2015packet}& 76.88 & 24.54 & 1.499 & NA\\
        \hline
         MPC \cite{mao2020robust}& 42.80 & 42.41 & 2.170 & $\times$\\
        \hline
	\end{tabular}\\
	\label{circlecomparetableperformance}
\end{table}



\subsection{Experiment 2: Eight-curve reference tracking}
\begin{figure*} 
    \centering
    \hfill
  \subfloat[\label{eig1a}]{%
       \includegraphics[width=0.3\linewidth]{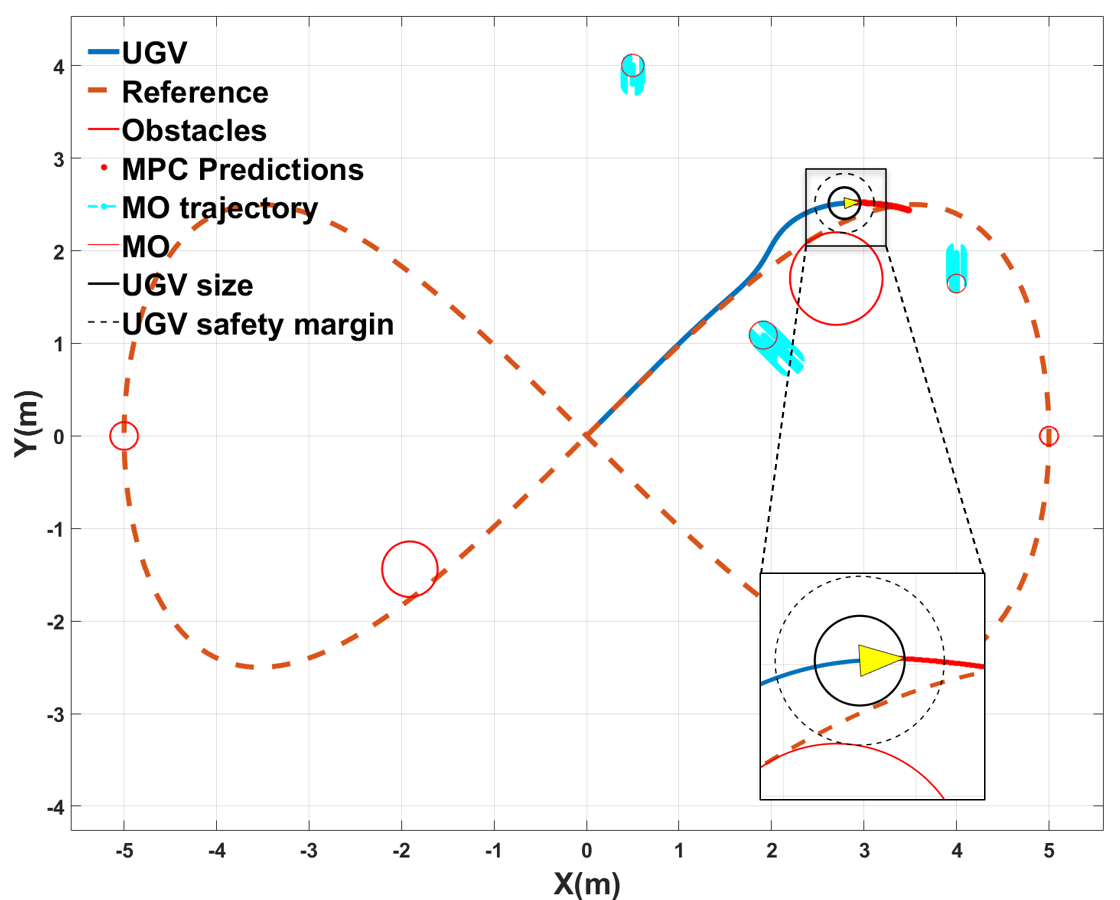}}
    \hfill
  \subfloat[\label{eig1b}]{%
        \includegraphics[width=0.3\linewidth]{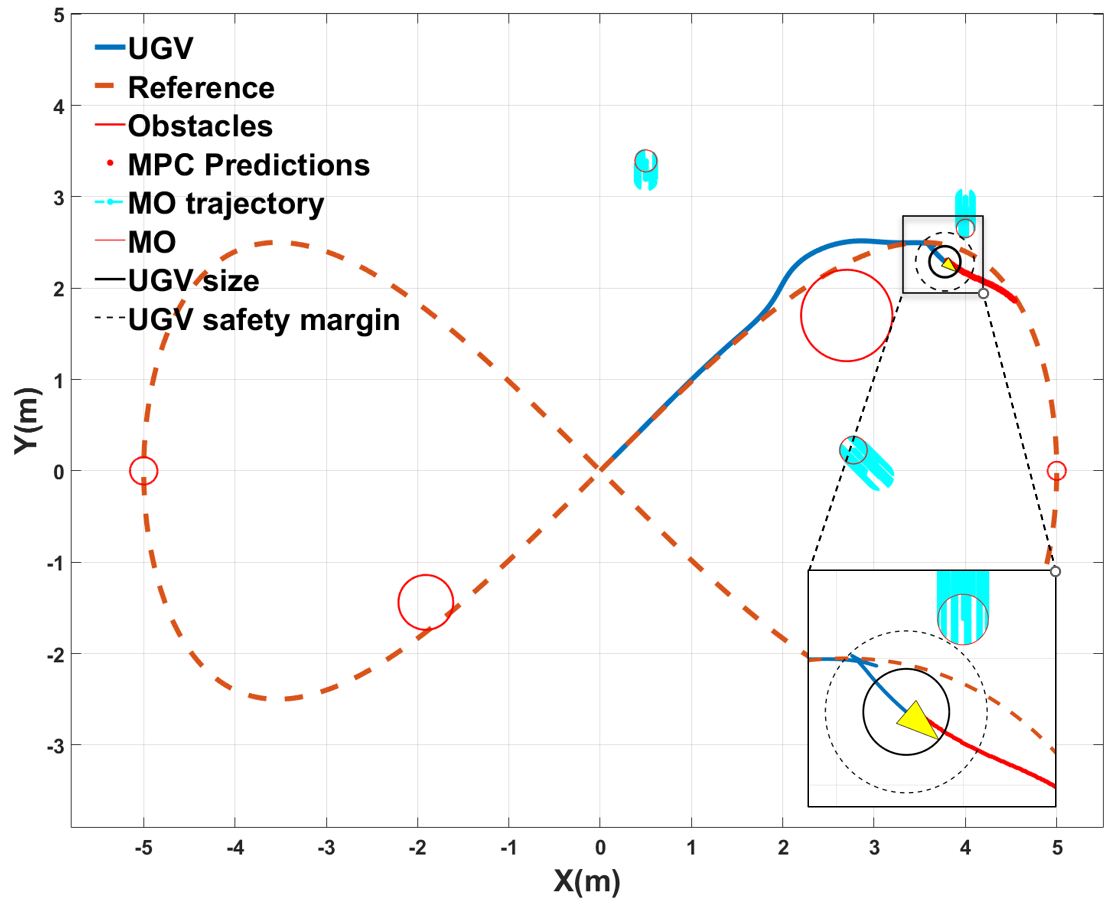}}
    \hfill
  \subfloat[\label{eig1c}]{%
        \includegraphics[width=0.3\linewidth]{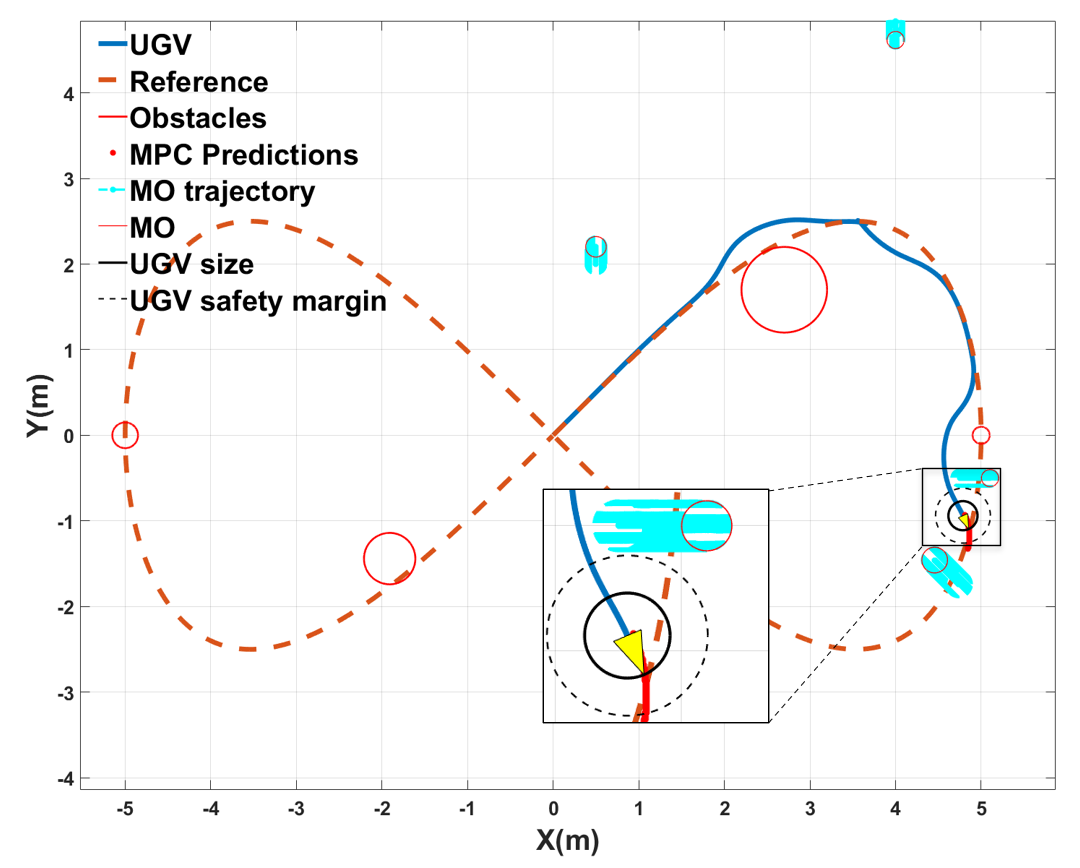}}
    \\
    \hfill
  \subfloat[\label{eig1d}]{%
        \includegraphics[width=0.3\linewidth]{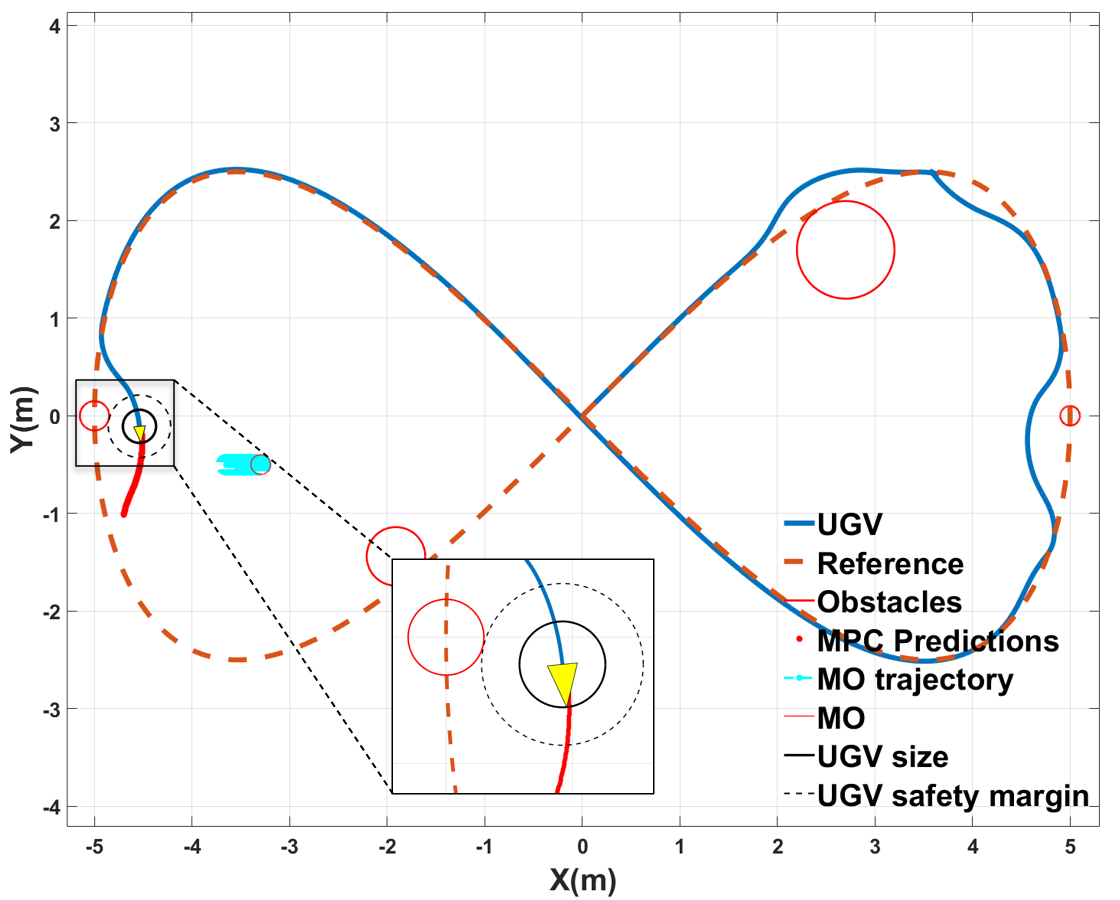}}
    \hfill
  \subfloat[\label{eig1e}]{%
        \includegraphics[width=0.3\linewidth]{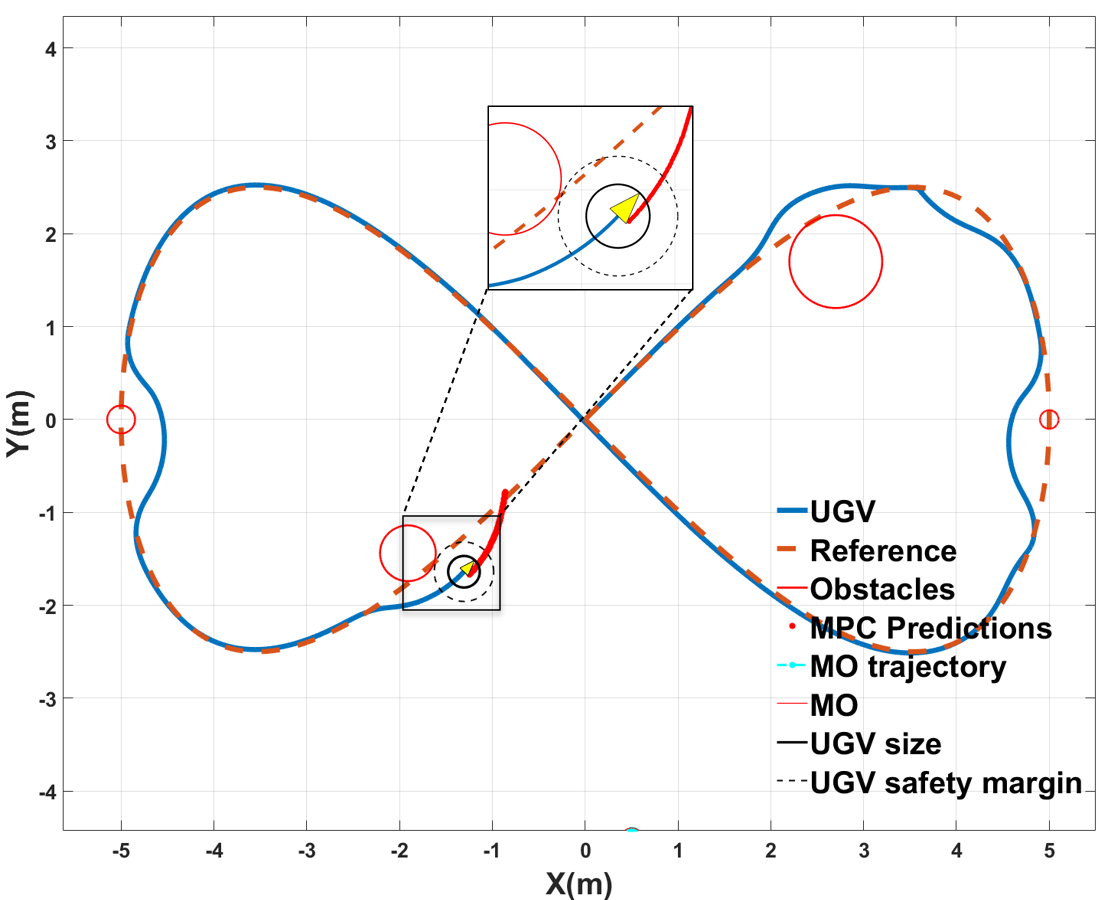}}
    \hfill
  \subfloat[\label{eig1f}]{%
        \includegraphics[width=0.3\linewidth]{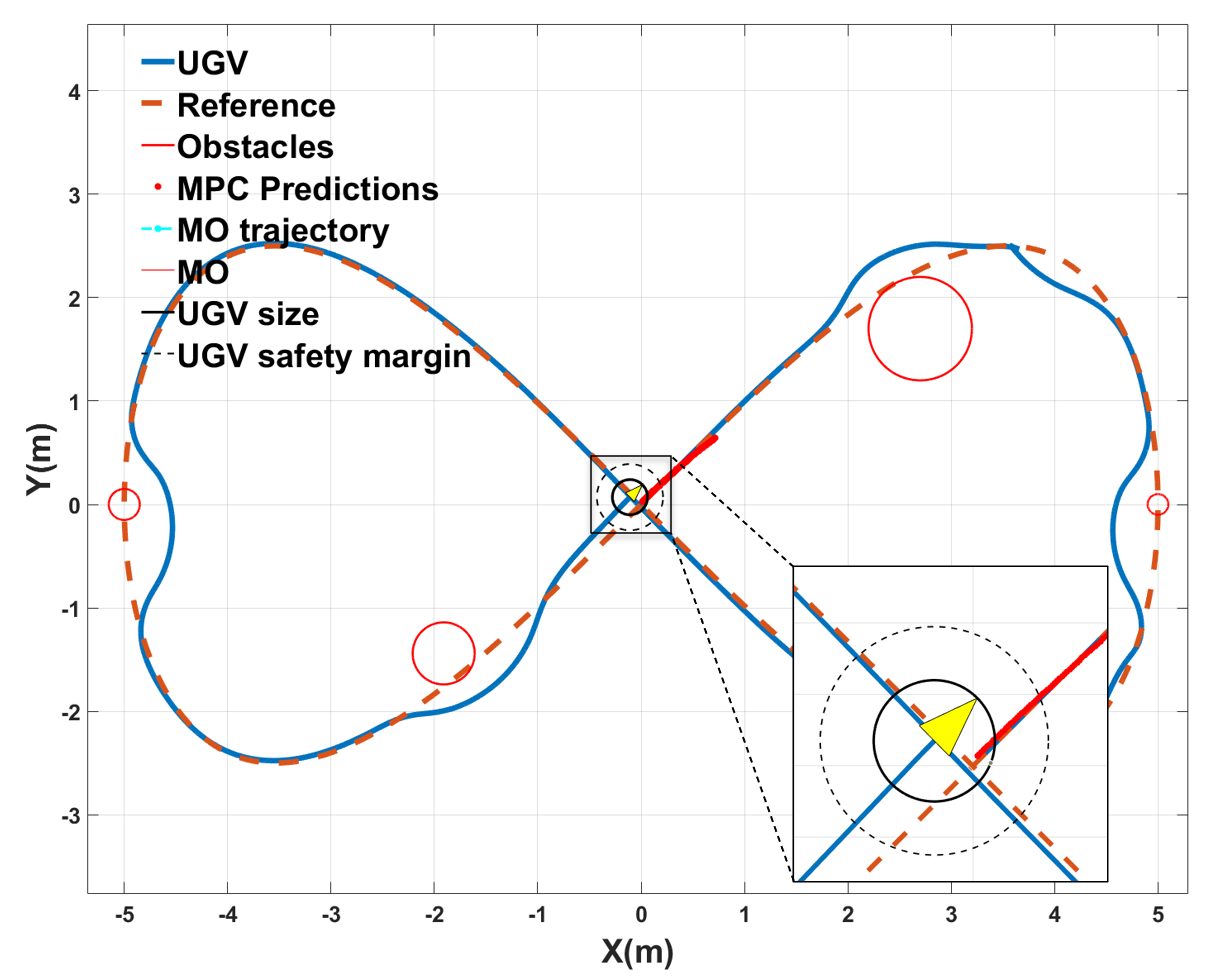}} 
  \caption{(a) Avoiding the rear of the first MO and the first SO. (b) Avoiding the second MO by predicting the motion of it. (c) Avoid the second SO and slow down to avoid collision with the third MO. (d), (e) Avoid the third and forth SO and get back to the reference track. (f) Safely reach the goal}
  \label{figeight} 
\end{figure*}

In this experiment, we conducted tests on our proposed scheme and other methods within a more complex scenario. Similar to the experiments we conducted before, the simulation encompassed various factors, including process noise, measurement noise, packet loss, delay, and obstacles. Instead of employing circular tracking, we utilized an eight-curve reference path, and the trajectory satisfies to the following mathematical relationships: 
\begin{align*}
    x_k^{ref} &= 5 \sin(\theta_k),\\
    y_k^{ref} &= 5 \sin(\theta_k) \cdot cos(\theta_k).
\end{align*}
Under the nominal experimental settings mentioned in section \ref{resultsanddiscussionoverview}, there are also typical collision avoidance actions shown in Figure \ref{figeight}. Like experiment 1, we include 4 MOs and 4 SOs in this simulation. The positions of the SOs are $(x^{so}_1,y^{so}_1) = (2.7, 1.7)$, $(x^{so}_2,y^{so}_2) = (5, 0)$, $(x^{so}_3,y^{so}_3) = (-1.9,1.4)$, and $(x^{so}_4,y^{so}_4) = (-5, 0)$, each with a respective radius of 0.5m, 0.1m, 0.3m, and 0.15m. And the starting states of moving obstacles are $(x^{mo}_1,y^{mo}_1,\theta^{mo}_1) = (0.5,2.5,\pi)$, $(x^{mo}_2,y^{mo}_2,\theta^{mo}_2) = (4,0,\pi/2)$, $(x^{mo}_3,y^{mo}_3,\theta^{mo}_3) = (-1.5,6,-\pi/2)$, and $(x^{mo}_4,y^{mo}_4,\theta^{mo}_4) = (6.5,3,-\pi)$ with moving speed of 0.25 m/s, 0.5 m/s, 0.2 m/s, 0.5 m/s, and radius 0.25 m, 0.15 m, 0.1 m and 0.15 m  respectively.

\begin{figure}[htbp]
\centerline{\includegraphics[width=0.4 \textwidth]{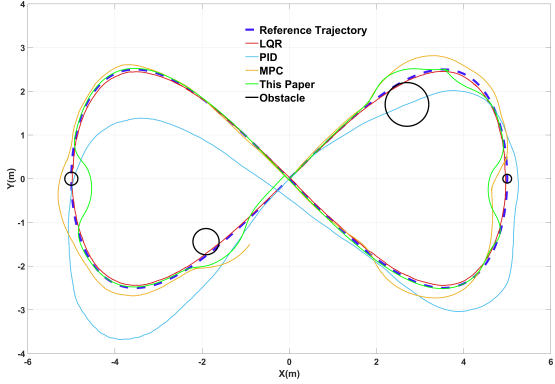}}
\caption{Experiment 2: Comparing of the performance of different methods}
\label{eightmpclqrpidcomparison}
\end{figure}




Every time our UGV try to avoid the obstacle, it will move off the track for a short period of time, and how big the detoured track would be depends on the placement and the size of the obstacles as shown in Figure \ref{figeight}. In Figure \ref{eig1a}, precise localization is achieved at this stage, and allowing the SO 1 edge of the barley approach to the safety margin and return to the eight-curve reference line after of avoiding action. Subsequently, as illustrated in Figure \ref{eig1b}, the UGV must navigate around MO 2 once it returns to its designated path. Figures \ref{eig1c} and \ref{eig1d} depict the UGV's ability to effectively avoid SO 2 and SO 3 while continuing along the eight trajectory. Following an extended period of obstacle-free travel, the UGV encounters the need to circumvent SO 3 and SO 4 towards the end of the journey. At this stage, the localization accuracy has diminished compared to earlier stages. Despite this, the UGV manages to complete the required trajectory, as demonstrated in Figure \ref{eig1f}. However, it may not be able to maintain a safe distance from all obstacles within the safety margin.

\begin{table}[t]
	\renewcommand\arraystretch{1}
	\centering
	\caption{Experiment 2: COMPARATIVE TABLE OF VARIOUS PAPERS AND METHODOLOGIES IN TERMS OF PERFORMANCE}
	\begin{tabular}{|p{1.2cm}|p{1.2cm}|p{1.2cm}|p{1.2cm}|p{1.2cm}|p{1.2cm}|p{1.2cm}|p{1.2cm}|p{1.2cm}|p{1.2cm}|}
		\hline
		Method & Multi-core Process Time (ms) & Average Euclidean distance error (cm) & Step Heading Error (deg) & Collision Avoidance\\
        \hline
        This paper & 33.92 & 13.48 & 0.405 & \checkmark\\
		\hline
         PID \cite{8104807}& 0.026 & 81.37 & 4.776 & NA\\
        \hline
         LQR \cite{nguyen2015packet}& 89.30 & 25.42 & 2.714 & NA\\
        \hline
         MPC \cite{mao2020robust}& 41.04 & 22.99 & 2.796 & $\times$\\
        \hline
	\end{tabular}\\
	\label{eightcomparetableperformance}
\end{table}
Fig \ref{eightmpclqrpidcomparison} illustrates the performance of different methods performing the eight-curve reference tracking task, including LQR from \cite{nguyen2015packet}, PID from \cite{8104807}, MPC from \cite{mao2020robust}. 
Table \ref{eightcomparetableperformance} shows a comparison that aligns with the findings from Experiment 1. In tasks of increased complexity, the proposed method demonstrates a more superior reliability over simpler tasks, evidenced by a markedly lower average Euclidean distance error when compared to other approaches. While LQR continues to exhibit commendable reference tracking control capabilities, its utility is limited by the high processing times. The performance of the PID controller is notably inadequate due to its vulnerability to uncertainties. To effectively operate in complex wireless environments, the PID controller necessitates substantial calibrations and adjustments. However, its significantly reduced processing time remains a distinctive advantage over competing methodologies.

\subsection{Experiment 3: Point stabilization}
\begin{figure*} 
    \centering
  \subfloat[\label{ps1a}]{%
       \includegraphics[width=0.3\linewidth]{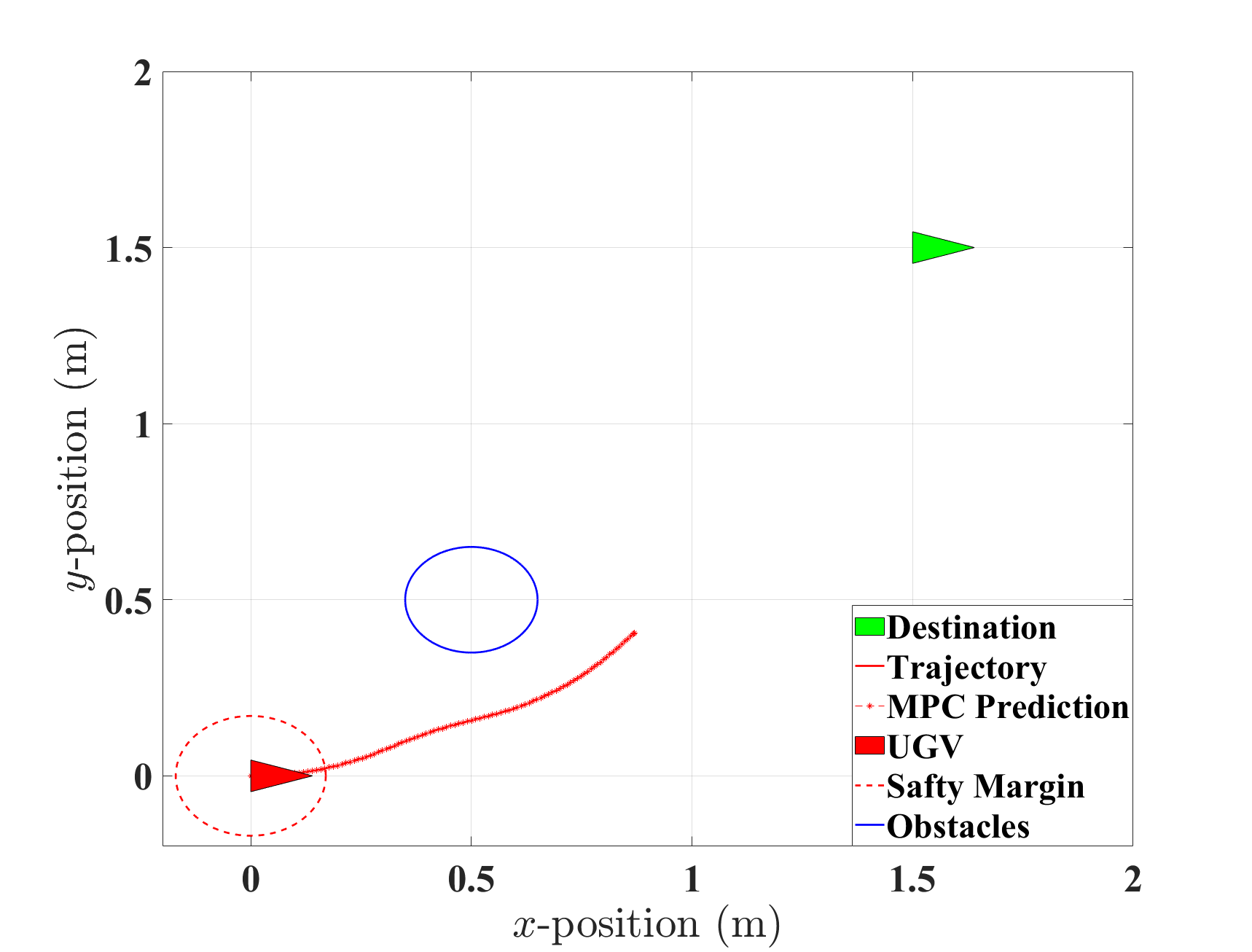}}
    \hfill
  \subfloat[\label{ps1b}]{%
        \includegraphics[width=0.3\linewidth]{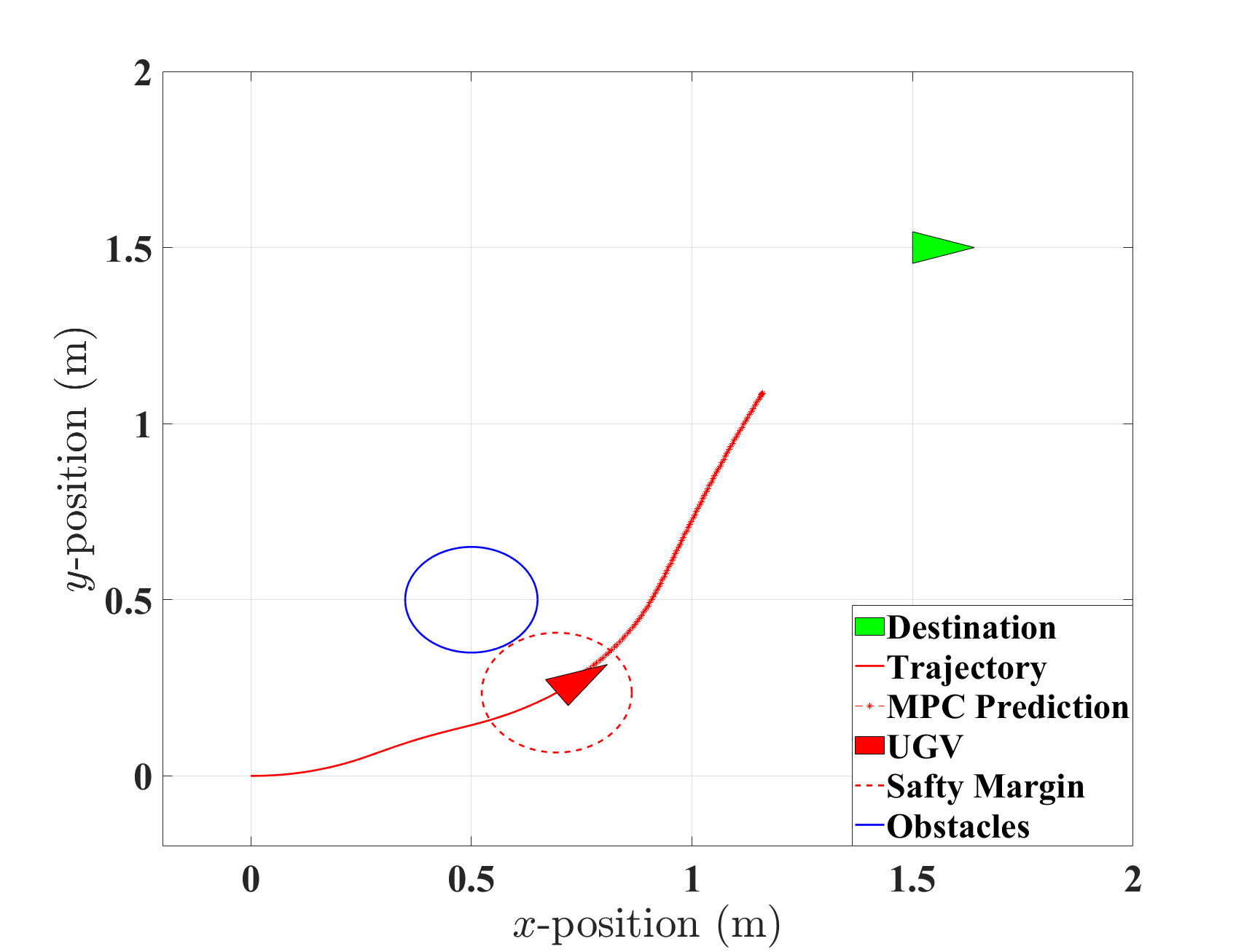}}
    \hfill
  \subfloat[\label{ps1c}]{%
        \includegraphics[width=0.3\linewidth]{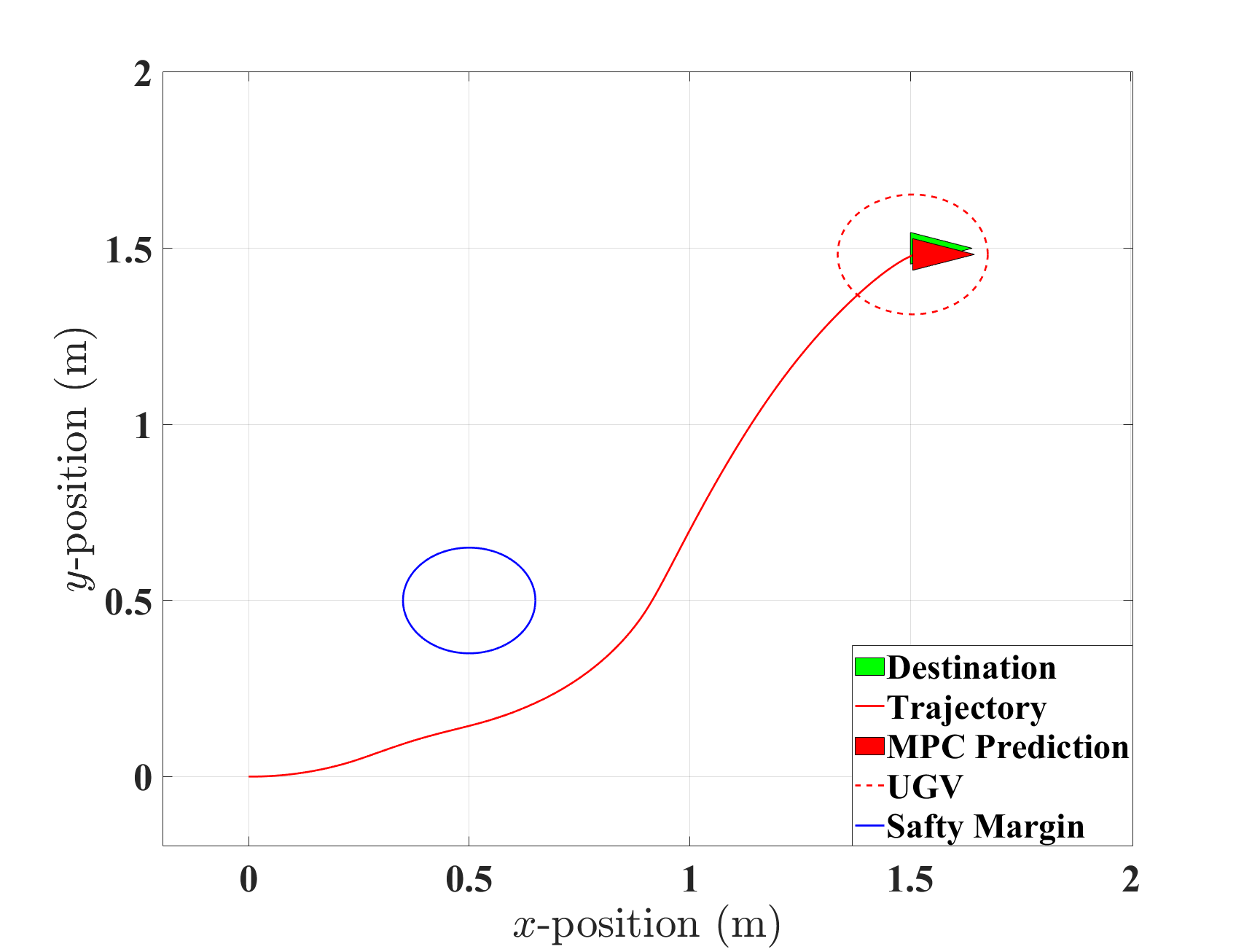}}
  \caption{(a) The UGV deployed at the origin. (b) The UGV follows the path lively generated by the MPC algorithm. (c) Almost perfectly algined with the destination state.}
  \label{figps} 
\end{figure*}
\begin{figure}[htbp]
\centerline{\includegraphics[width=0.4 \textwidth]{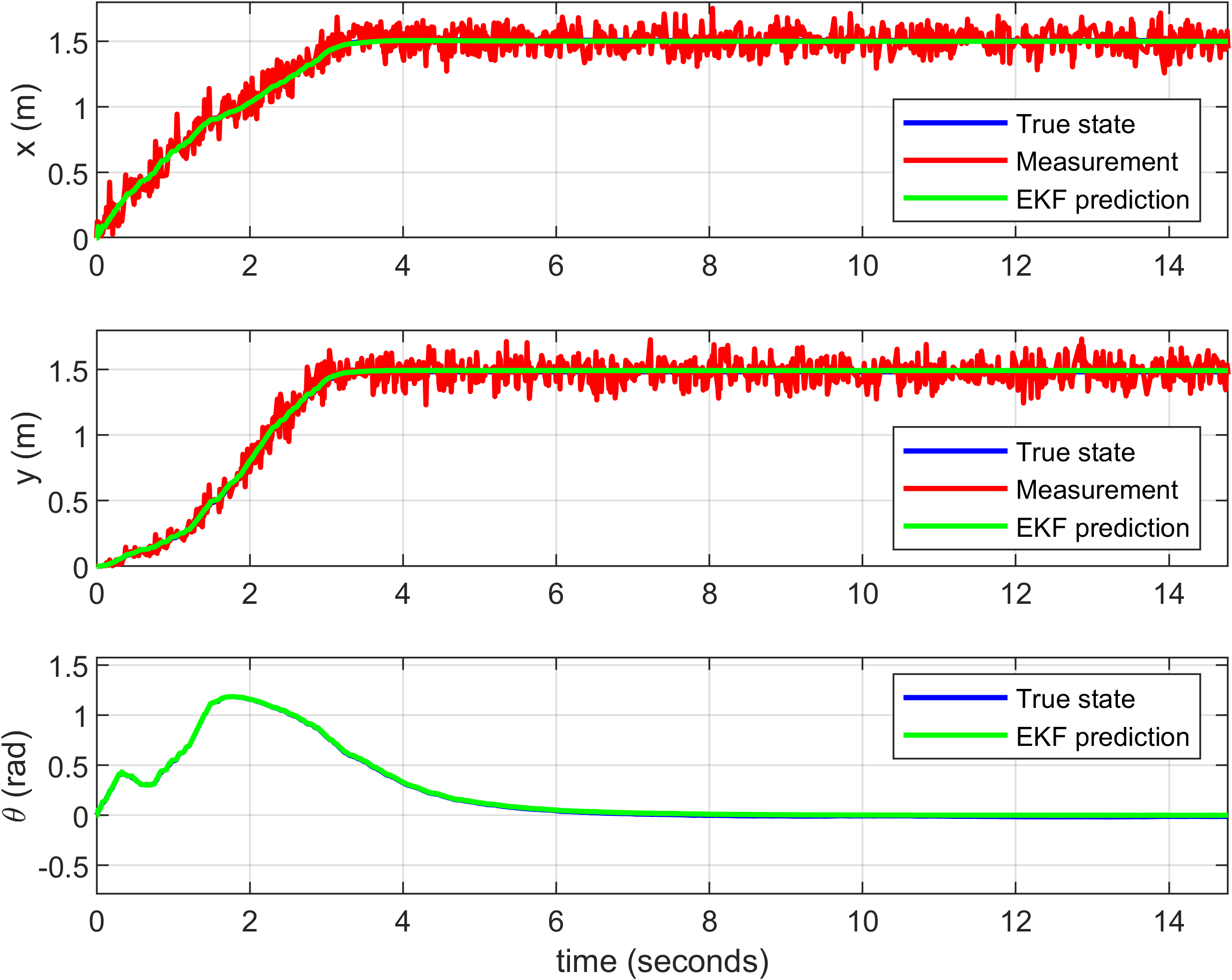}}
\caption{Experiment 3: Actual, measured and estimated states of the UGV}
\label{ps}
\end{figure}

In this experiment, we evaluated the efficacy of our proposed control strategy through a basic point stabilization challenge. The robot was initialized at a position given by $[x_i, y_i, \theta_i] = [0, 0, 0]$. The target position, denoted by $s_{ref}$, was set to $[x_g, y_g, \theta_g] = [1.5, 1.5, 0]$. Additionally, a circular obstacle with a diameter of 0.3 meters was positioned at the center coordinates $[x_o, y_o] = [0.5, 0.5]$. Additionally,  we take into account process noise, measurement noise, packet loss, and delay. Figure \ref{ps} shows the $x,y$ and $\theta$ states of our proposed method under the nominal experimental settings. The steady-state error is $1.82$ centimeters, and the heading error is $0.0475$ in radians.
The results demonstrate that the proposed scheme exhibits satisfactory performance even in the presence of packet loss, communication latency, process noise, and measurement noise.

\begin{figure}[htbp]
\centerline{\includegraphics[width=0.4 \textwidth]{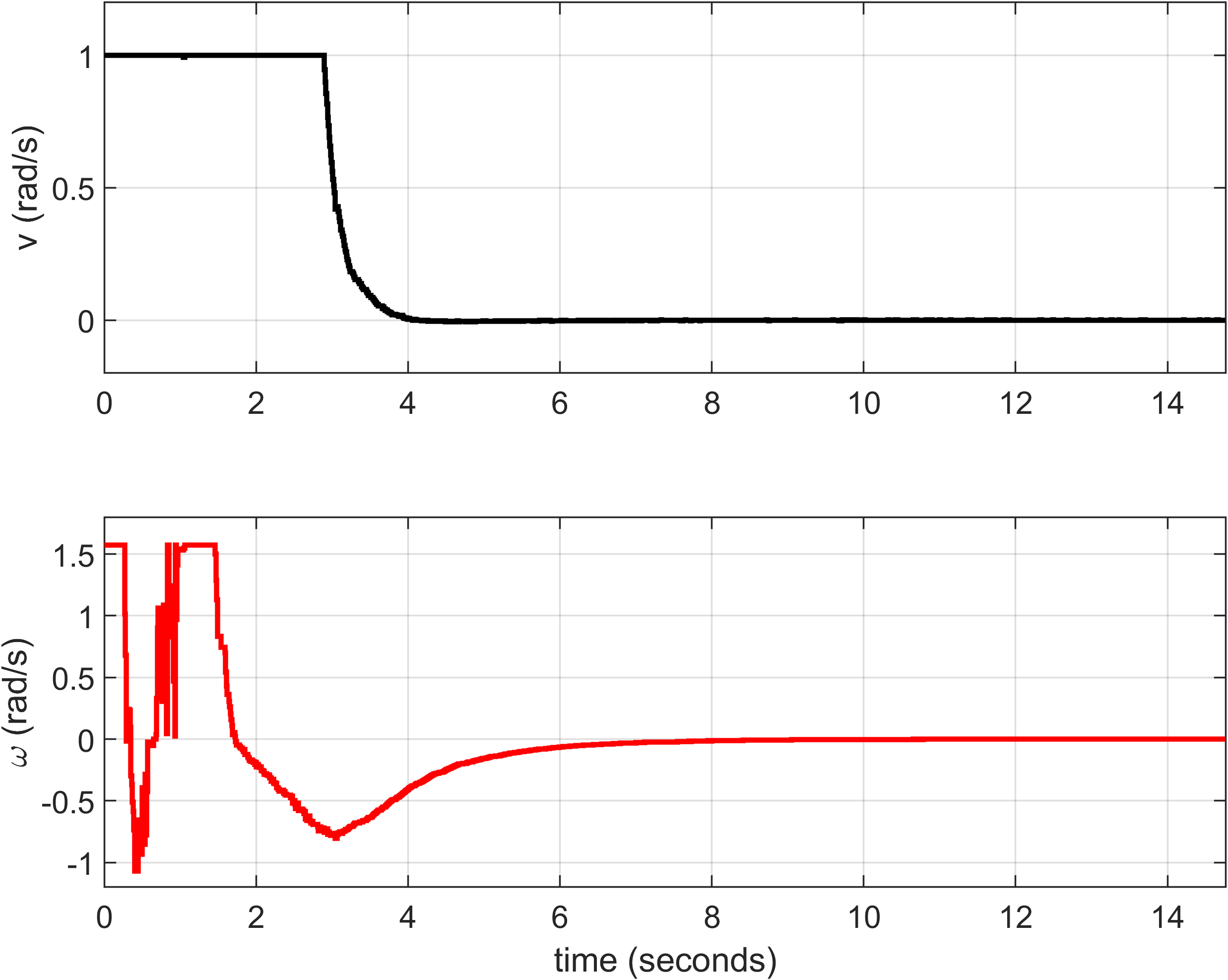}}
\caption{Experiment 3: Longitudinal velocity and yaw rate applied to the UGV}
\label{psvw}
\end{figure}

To illustrate the obstacle avoidance performance, we could see significant variations in angular velocity ($\omega$) shown in Figure \ref{psvw} is due to the obstacle avoidance actions. As depicted in Figure \ref{figps}, the trajectory of the UGV exhibits a pronounced curvature to bypass the obstacle successfully. Under nominal simulation conditions, the accomplishment of this task can be ensured without compromising safety.

The performance of the proposed control strategy is associated with the length of the prediction horizon of the MPC. However, a longer MPC horizon does not necessarily guarantee improved performance; as the prediction horizon increases, the MPC processing time also increases. Under the nominal communication channel conditions, Table \ref{psempctime} illustrates the average processing time of the MPC for each iteration with varying prediction horizons.
\begin{table}[t]
	\renewcommand\arraystretch{1}
	\centering
	\caption{EXPERIMENT 3: SINGLE CORE AND MULTI-CORE MPC TIME COMPARISON FOR POINT STABILIZATION PROBLEM}
	\begin{tabular}{|p{1.2cm}|p{1.2cm}|p{1.2cm}|p{1.2cm}|p{1.2cm}|}
		\hline
		Prediction horizon & Single core MPC time (ms) & Multi-core MPC time (ms) & Steady state position error (cm) & Steady state heading error (deg) \\
        \hline
        10 & 12.0 & 9.7 & 29.37 & 35.02\\
        \hline
        15 & 11.3 & 9.2 & 5.33 & 21.45\\
        \hline
        20 & 10.6 & 8.4 & 16.4 & 7.341\\
        \hline
        50 & 10.8 & 8.7 & 16.5 & 0.5910\\
        \hline
        100 & 14.4 & 11.6 & 14.7 & 0.3617\\
		\hline
	\end{tabular}\\
	\label{psempctime}
\end{table}

The results of Table \ref{psempctime} reveals that the increase of prediction horizons in the MPC will result lower steady state position error as well as heading error, however, increasing the prediction horizon will also cause a longer processing time. As the prediction horizon decreases, the steady-state errors rise considerably. This is attributed to the difficulty faced by the controller in forecasting future motion based on limited available information when the prediction horizon is extremely short.

To assess the efficacy of our proposed scheme, we conducted a comparative analysis of its performance across various communication channel conditions. Table \ref{pssnr} presents the steady-state errors observed under differing wireless network communication specifications.
The results indicate that for a relatively straightforward task such as point stabilization, our proposed scheme consistently delivers strong performance across diverse communication network specifications.
\begin{table}[t]
	\renewcommand\arraystretch{1}
	\centering
	\caption{EXPERIMENT 3: COMMUNICATION CHANNEL STATUS COMPARISON FOR POINT STABILIZATION TASK}
	\begin{tabular}{|p{1cm}|p{1cm}|p{1cm}|p{1cm}|p{1cm}|p{1.2cm}|}
		\hline
		SNR & S-to-C maximum latency (ms)& S-to-C Packet error rate & Steady state position error (cm) & Steady state heading error (deg) & Collision Avoidance  \\
        \hline
        5 &10& 0.696 & 1.36 & 0.3275 & \checkmark\\
        \hline
        5 &30& 0.696 & 1.44 & 0.0495 & \checkmark\\
        \hline
        5 &50& 0.696 & 1.89 & 0.7269 & \checkmark\\      
		\hline
        10 &10& 0.312 & 1.26 & 0.172 & \checkmark\\
		\hline
        10 &30& 0.312 & 1.70 & 0.2181 & \checkmark\\
		\hline
        10 &50& 0.312 & 1.09 & 0.2619 & \checkmark\\
		\hline
        15 &10& 0.112 & 1.65 & 0.1151 & \checkmark\\
		\hline
        15 &30& 0.112 & 1.38 & 0.4905 & \checkmark\\
		\hline
        15 &50& 0.112 & 1.82 & 0.0475 & \checkmark\\
		\hline
        20 &10& 0.037 & 1.26 & 0.1688 & \checkmark\\
		\hline
        20 &30& 0.037 & 1.60 & 0.7086 & \checkmark\\
		\hline
        20 &50& 0.037 & 1.18 & 0.5819 & \checkmark\\
		\hline
        100 &10& 0 & 1.60 & 0.8037 & \checkmark\\
		\hline
        100 &30& 0 & 1.37 & 0.0411 & \checkmark\\
		\hline
        100 &50& 0 & 1.24 & 0.1117 & \checkmark\\
		\hline
	\end{tabular}\\
	\label{pssnr}
\end{table}

\section{Comparison}

\begin{figure*} 
    \centering
    \subfloat[\label{fig:boxa}]{%
        \includegraphics[width=0.3\linewidth]{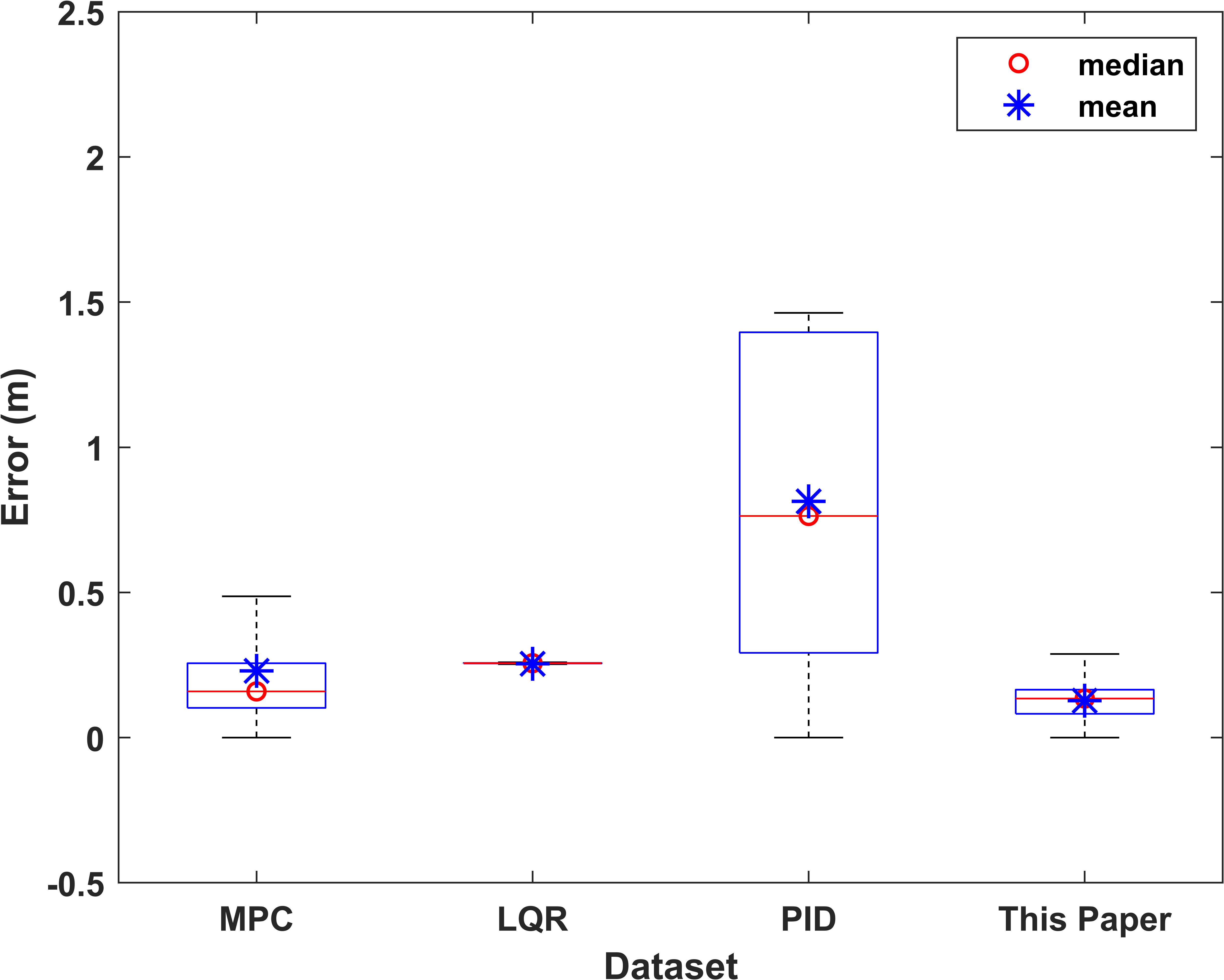}}
    \subfloat[\label{fig:boxb}]{%
        \includegraphics[width=0.3\linewidth]{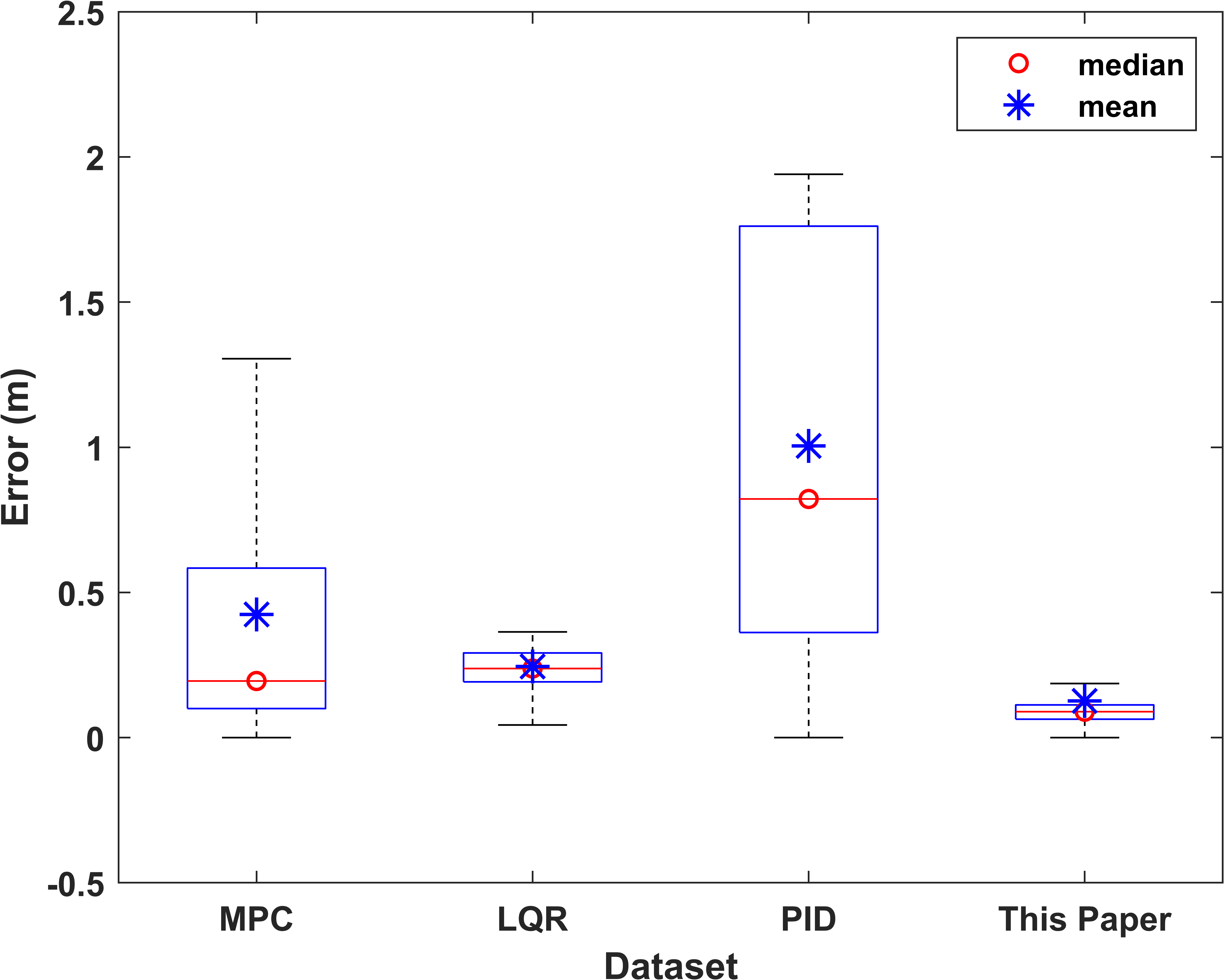}}
    
    \subfloat[\label{fig:boxc}]{%
        \includegraphics[width=0.3\linewidth]{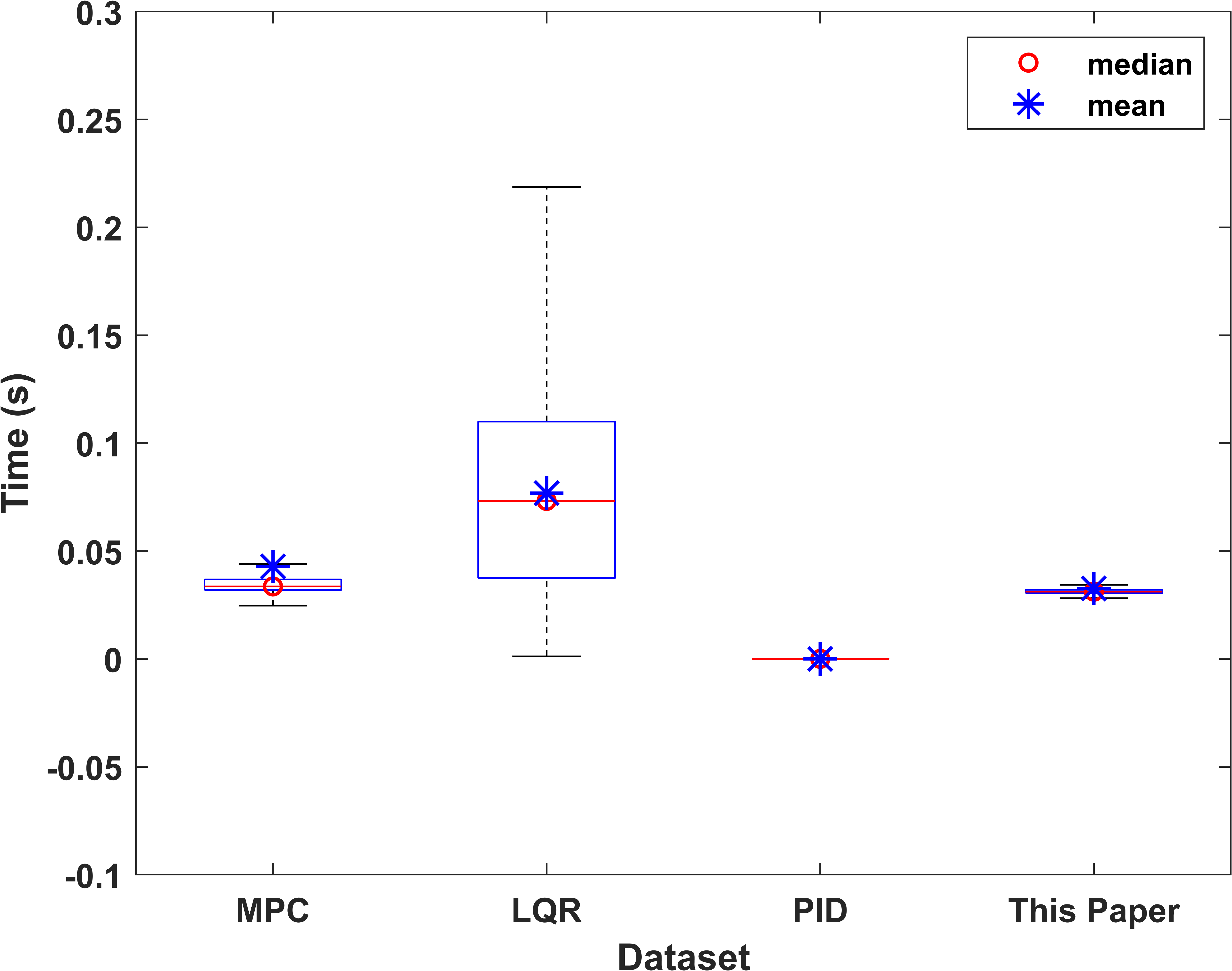}}
    \subfloat[\label{fig:boxd}]{%
        \includegraphics[width=0.3\linewidth]{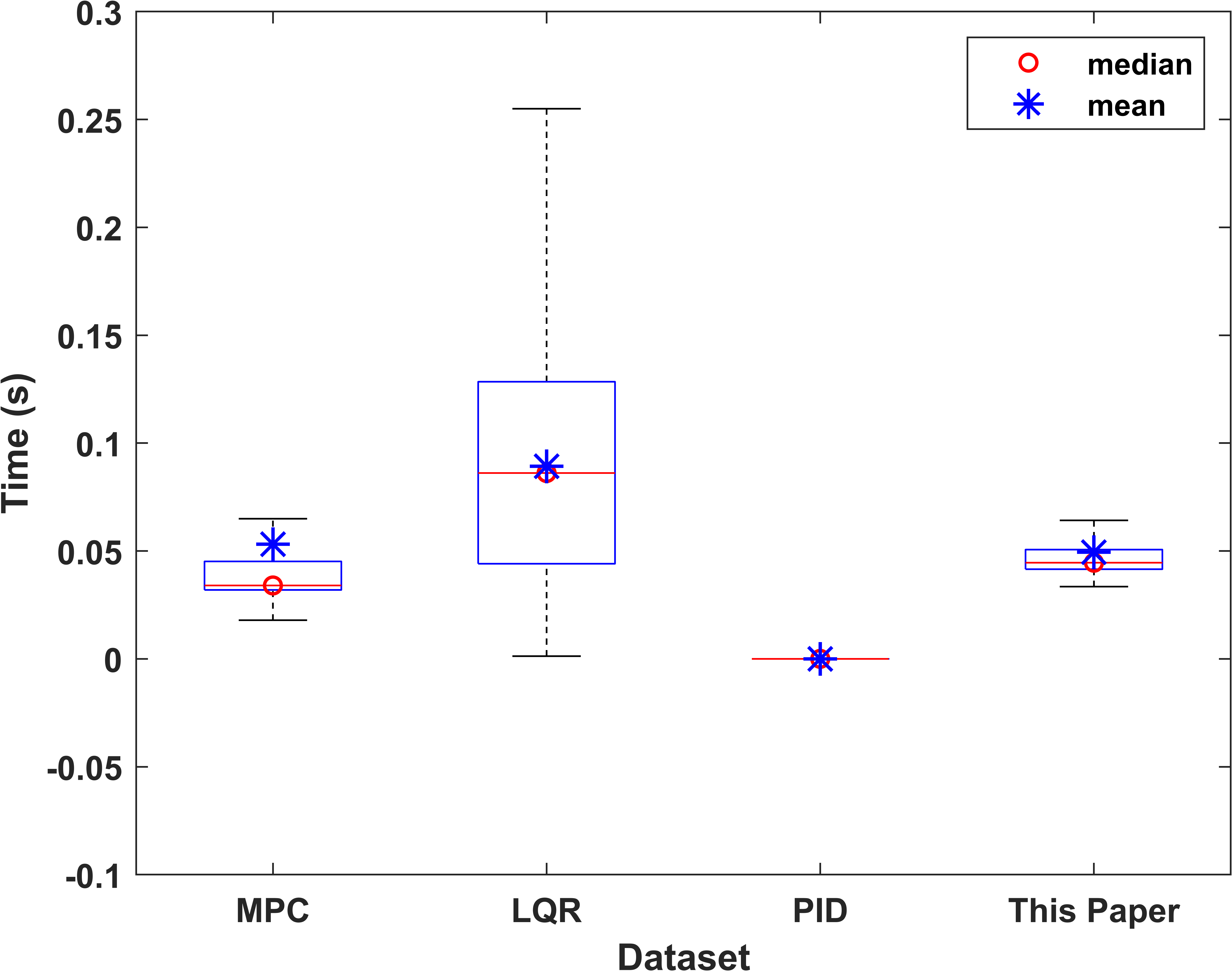}}

    \caption{(a) Circular reference tracking Euclidean distance error box plot (b) Eight-curve reference tracking Euclidean distance error box plot (c) Circular reference tracking processing time box plot (d) Eight-curve reference tracking processing time box plot }
    \label{4boxplots} 
\end{figure*}

In comparing with other methodologies, here we have used four different methods including the one that mentioned in this paper to solve the same type of problem (circular-curve reference tracking problem and eight-curve reference tracking problem).

Figure \ref{4boxplots} illustrates that the method proposed by this study excels over other techniques in terms of Euclidean distance error. Figure \ref{fig:boxa} and \ref{fig:boxb} detail the Euclidean distance error comparisons for Experiment 1 and 2, respectively. The outcomes from both experiments underscore that the proposed method consistently registers the smallest Euclidean distance errors when compared with existing methods. Despite LQR's exceptional consistency in circular reference tracking tasks without collision avoidance, its average distance error is still surpassed by our proposed method. Contrary to the pronounced variability in Euclidean distance errors, the average processing times depicted in Figure \ref{fig:boxc} and \ref{fig:boxd} highlight PID's efficiency. The processing overhead of PID is minimal among all methods reviewed, whereas LQR experiences considerably longer processing times in comparison.

In summary, even with extended routes due to obstacle avoidance strategies, the proposed method in this paper retains optimal tracking accuracy in Euclidean distance. Regarding processing time, our method demonstrates effective performance without demanding significant computational power.

\section{Conclusion}
\label{conclusion}
In this paper, we present a wireless networked control scheme that takes into account the challenges posed by an imperfect communication environment, such as packet loss and delays. By solving the multiple-shooting MPC problem, our control scheme could provide a validate solution for WNCS problems. Additionally, we employ the EFK to attenuate both process and measurement noise. This control scheme is applied to address diverse tasks, including point stabilization, circular curve tracking control, and eight-curve tracking control. In these experiments, we rigorously evaluate the reference tracking accuracy and the performance of the obstacle avoidance capability of our proposed method when implemented on a UGV. The results demonstrate the efficacy of our proposed scheme in delivering satisfactory performance for these tasks.

Regarding future research directions, we aspire to extend the application of this method to more complex scenarios. For instance, we aim to integrate this method with pathfinding algorithms to tackle more intricate tasks within environments that closely simulate real-world conditions.

\bibliographystyle{unsrt}
\bibliography{citation.bib}
\vfill

\begin{IEEEbiography}[{\includegraphics[width=1in,height=1.25in,clip,keepaspectratio]{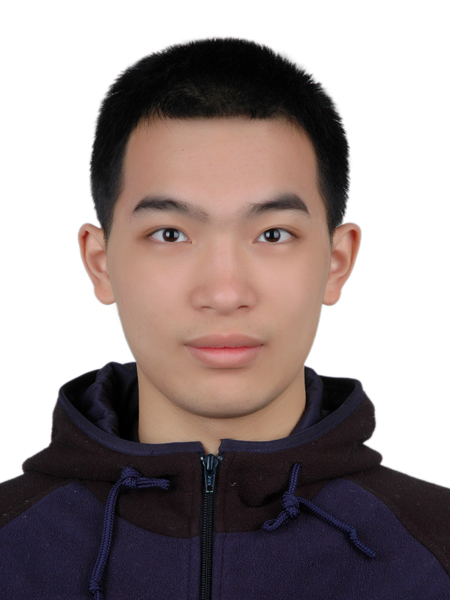}}]{Jinghao Cao} received the BSc in Electrical Engineering Systems and MEng in Electrical and Electronic Engineering from the University of Melbourne, Parkville campus in 2020 and 2022. He is currently a PhD Student at the School of Electrical and Information Engineering, The University of Sydney. His research interests includes wireless network control systems (WNCS), industrial Internet of Things (IIoT), and robotic control systems.
\end{IEEEbiography}

\begin{IEEEbiography}[{\includegraphics[width=1in,height=1.25in,clip,keepaspectratio]{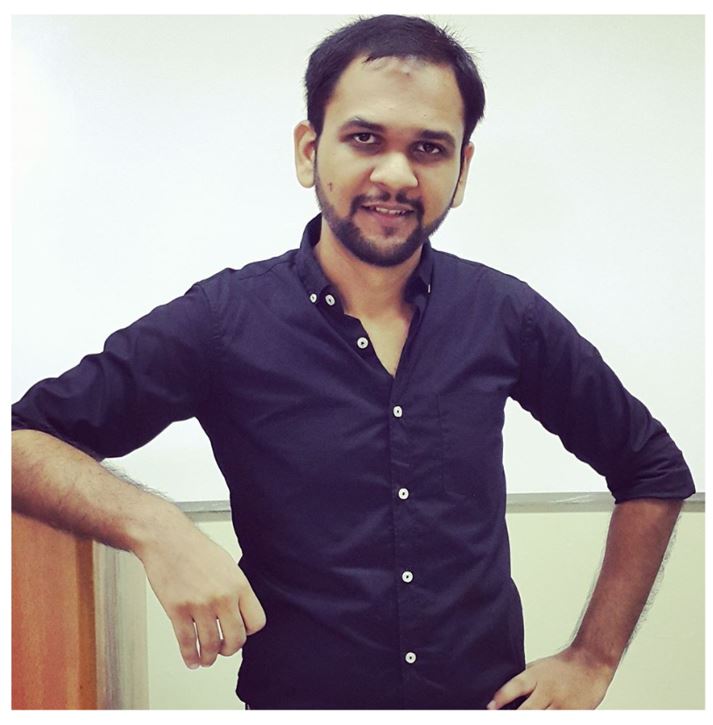}}]{Dr. Subhan Khan} received the Ph.D. degree in Mechatronic Engineering from the School of Mechanical and Manufacturing Engineering, University of New South Wales (UNSW), Sydney in 2022. He received his BSc in Computer Engineering and MSc in Electronic Communication and Computer Engineering degrees from COMSATS Institute of Information Technology (CIIT) Lahore, Pakistan and the University of Nottingham, UK campus, in 2013 and 2015, respectively. His research interests include perception, planning and control of robotic systems. He was the recipient of the university international postgraduate award (UIPA) 2018-2021 and the campus silver medalist in 2013.  Currently, he is working as a Postdoctoral Research Associate in robotics and automation engineering at the University of Sydney.
\end{IEEEbiography}
\begin{IEEEbiography}[{\includegraphics[width=1in,height=1.25in,clip,keepaspectratio]{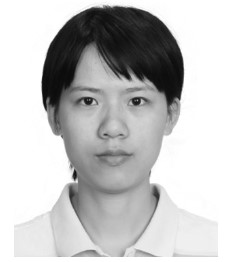}}]{Dr. Wanchun Liu} (Member, IEEE) received the B.S. and M.S.E. degrees in electronics and information engineering from Beihang University, Beijing, China, and Ph.D. from the Australian National University, Canberra, Australia. Following her graduation, she joined the University of Sydney as a research fellow. Her research interest lies in communications and networked control theory, wireless control for industrial Internet of Things (IIoT), and wireless human-machine collaborations for Industry 5.0. She was a co-chair of the Australian Communication Theory Workshop in 2020-2021. She was a recipient of the Australian Research Council’s Discovery Early Career Researcher Award 2023, the Dean’s Award for Outstanding Research of an Early Career Researcher 2022, and the Chinese Government Award for Outstanding Students Abroad 2017.
\end{IEEEbiography}
\begin{IEEEbiography}[{\includegraphics[width=1in,height=1.25in,clip,keepaspectratio]{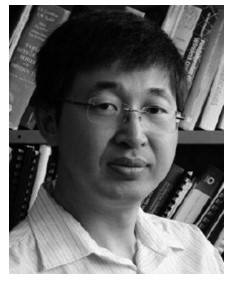}}]{Prof. Yonghui Li} (Fellow, IEEE) received the Ph.D. degree from Beijing University of Aeronautics and Astronautics, Beijing, China, in November 2002.
Since 2003, he has been with the Centre of Excellence in Telecommunications, The University of Sydney, Sydney, NSW, Australia, where he is currently a Professor and the Director of the Wireless Engineering Laboratory, School of Electrical and Information Engineering. His current research interests are in the area of wireless communications, with a particular focus on MIMO, millimeter-wave communications, machine-to-machine communications, coding techniques, and cooperative communications. He holds a number of patents granted and pending in these fields.
Prof. Li is the recipient of the Australian Queen Elizabeth II Fellowship in 2008 and the Australian Future Fellowship in 2012. He received the Best Paper Awards from the IEEE International Conference on Communications 2014, IEEE PIRMC 2017, and IEEE Wireless Days Conferences 2014. He is currently an Editor of the IEEE Transactions on Communications and IEEE Transactions on Vehicular Technology. He also served as the Guest Editor for several IEEE journals, such as IEEE Journal on Selected Areas in Communications, IEEE Communications Magazine, IEEE Internet of Things Journal, and IEEE Access.
\end{IEEEbiography}
\begin{IEEEbiography}[{\includegraphics[width=1in,height=1.25in,clip,keepaspectratio]{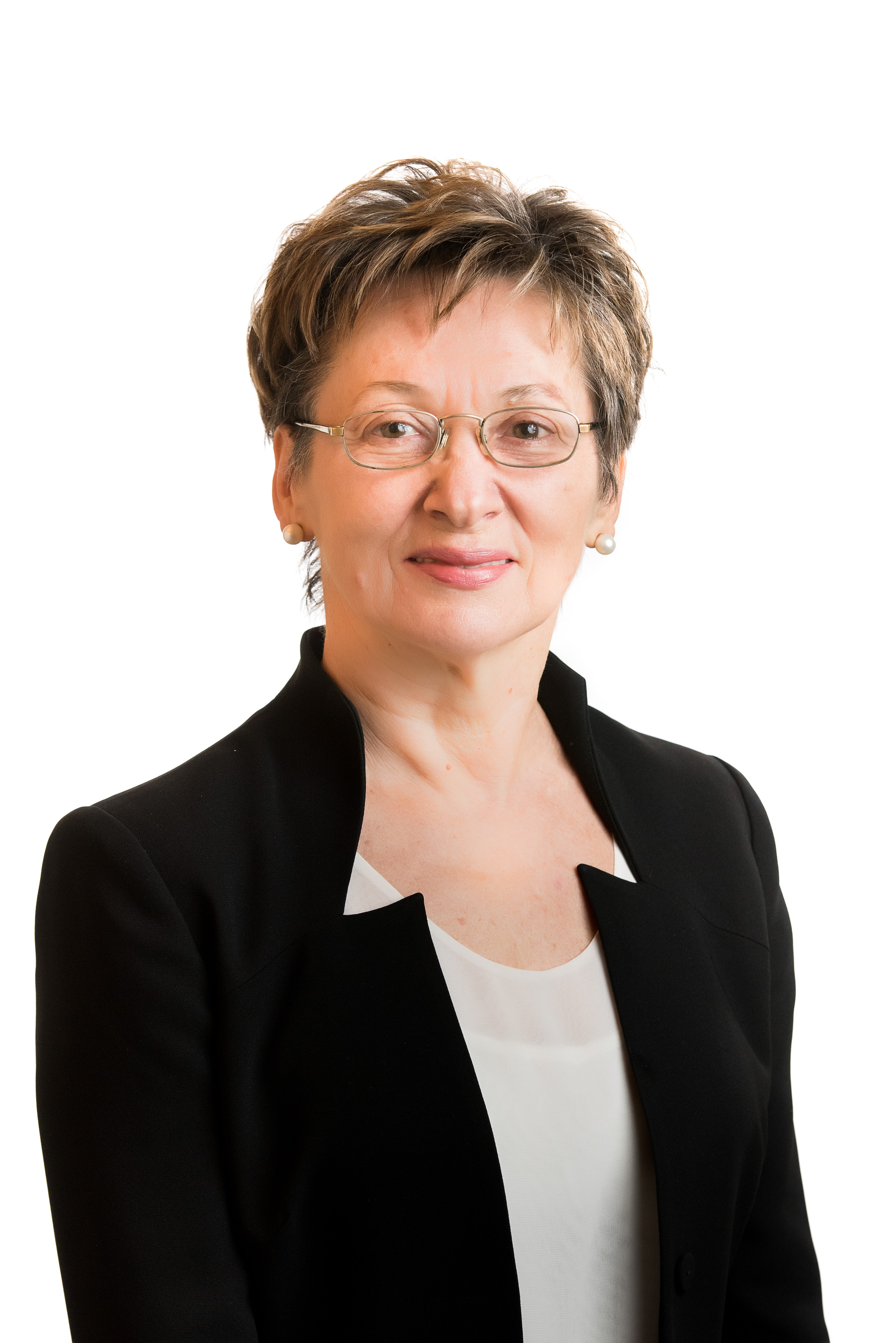}}]{Prof. Branka Vucetic} (Fellow, IEEE)  is an Australian
Laureate Fellow, a Professor of Telecommunications, and Director of the Centre for IoT and
Telecommunications at the University of Sydney.
Her current research work is in wireless networks,
and Industry 5.0. In the area of wireless networks,
she works on communication system design for 6G
and wireless AI. In the area of Industry 5.0, Vucetic’s
research is focused on the design of cyber-physicalhuman systems and wireless networks for applications in healthcare, energy grids, and advanced
manufacturing. Branka Vucetic is a Fellow of IEEE, the Australian Academy
of Technological Sciences and Engineering and the Australian Academy of
Science.
\end{IEEEbiography}

\end{document}